\documentclass[amsmath, amsfonts,twocolumn, superscriptaddress,floatfix,aps]{revtex4}

\usepackage{graphicx}
\graphicspath{{images/}}
\usepackage{lipsum}
\usepackage{bm}
\usepackage{dsfont}
\usepackage{amssymb}
\usepackage[dvipsnames]{xcolor}
\usepackage{pstricks}
\usepackage{verbatim}
\usepackage{units}
\usepackage{natbib}
\usepackage{mathtools}
\usepackage{multirow}
\usepackage{enumerate}
\usepackage{mathrsfs}
\usepackage{leftidx}
\usepackage{xspace}
\usepackage{amsmath}
\usepackage[normalem]{ulem} 
\usepackage{soul}
\usepackage{physics}
\usepackage{tabularx}
\usepackage{upgreek}
\usepackage[normalem]{ulem} 
\usepackage{comment} 
\newcolumntype{Y}{>{\centering\arraybackslash}X}

\usepackage{hyperref}
\hypersetup{colorlinks=true,linktoc=all,linkcolor=blue,breaklinks=true,citecolor=blue,urlcolor=blue}

\usepackage{braket}

\usepackage[english]{babel}
\addto\captionsenglish{}

\makeatletter
\renewcommand\paragraph{\@startsection{paragraph}{4}{\z@}%
  {3.25ex \@plus1ex \@minus.2ex}%
  {-0em}%
  {\normalfont\normalsize\itshape\indent}}
\begin{document}

\title{
Thermalization in Spatially Extended Open Quantum Systems:\\ Local versus Global Markovian Evolution}

\author{Jorge Tabanera-Bravo}
\email{jtabane@mpinat.mpg.de}
\affiliation{Mathematical bioPhysics group, Max Planck Institute for Multidisciplinary Sciences, Göttingen 37077, Germany}
\affiliation{Departamento de Estructura de la Materia, F\'isica T\'ermica y Electr\'onica and GISC, Universidad Complutense de Madrid, Pl. de las Ciencias 1. 28040 Madrid, Spain}

\author{Massimiliano Esposito}
\email{massimiliano.esposito@uni.lu}
\affiliation{Complex Systems and Statistical Mechanics, Department of Physics and Materials Science, University of Luxembourg, 30 Avenue des Hauts-Fourneaux, L-4362 Esch-sur-Alzette, Luxembourg}

\author{Felipe Barra}
\email{fbarra@dfi.uchile.cl }
\affiliation{Departamento de F\'isica, Facultad de Ciencias F\'isicas y Matem\'aticas, Universidad de Chile, 837.0415 Santiago, Chile}

\author{Juan M. R. Parrondo}
\email{parrondo@ucm.es}
\affiliation{Departamento de Estructura de la Materia, F\'isica T\'ermica y Electr\'onica and GISC, Universidad Complutense de Madrid, Pl. de las Ciencias 1. 28040 Madrid, Spain}

\date{\today}

\begin{abstract}
We investigate the dynamics of a qubit chain locally coupled to a thermal reservoir, modeled through repeated collisions with particles drawn from a heat bath. 
Under suitable conditions, the resulting Lindblad equation is thermodynamically consistent---it drives the system toward thermal equilibrium---while remaining local at short times. 
This framework reveals a crossover between the global Lindblad equation derived from the secular approximation in weak-coupling theory and the local dissipative models often employed in the literature, which generally fail to ensure thermodynamic consistency.
\end{abstract}

\maketitle
\setlength{\tabcolsep}{4pt}
\renewcommand{\arraystretch}{1.3}


\section{Introduction}

Understanding and controlling the dynamics of open quantum systems is central to both quantum thermodynamics and emerging quantum technologies, and the topic has a long and rich history~\cite{Van_Hove1954-ta, Fano1957-ob, Redfield1957-za, Redfield1965-gm, Haake1973-eu, Gorini1976-iq, Lindblad1976-uz}. 
A cornerstone of this field is the weak system--bath coupling regime, where a perturbative expansion combined with the secular approximation yields a Lindblad quantum master equation that guarantees thermalization when the system interacts with a single heat bath~\cite{Davies1974-ho, davis1976, SpohnLebowitz1978, Spohn1980-vb}. 
This equation captures both the decay of quantum coherences in the eigenbasis of the system Hamiltonian and the population dynamics among eigenstates, with transition rates obeying detailed balance. 
The combined action of these mechanisms drives the system toward a Gibbs state at the bath temperature. 
Modern and pedagogical treatments can be found in Refs.~\cite{CohenTBook, breuer2007, rivas2012, Strasberg2022-lu}.

Despite its success, the weak-coupling approach raises subtle questions when applied to spatially extended systems composed of multiple coupled, identical subsystems (e.g., tight-binding chains or spin arrays) that interact locally with a thermal bath (e.g., through a single subsystem)~\cite{Levy_2014, DeChiara2018Nov, RevModPhys.94.045006, Cattaneo_2019, potts_thermodynamically_2021,chenEfficientQuantumThermal2025}. 
The secular approximation relies on a clear separation of timescales: the characteristic oscillation periods associated with the energy differences between system eigenstates must be much shorter than the relaxation time induced by the bath. 
However, when the coupling between subsystems becomes weak, many of these energy gaps approach zero, and the corresponding oscillations slow down, thereby invalidating the secular approximation. 
In this limit, the subsystem directly connected to the bath effectively decouples from the rest of the system, and one expects a form of dynamical locality to emerge in the master equation, an aspect that the global weak-coupling formalism fails to capture.

To address this issue, we investigate a chain of qubits coupled to a thermal bath through repeated collisions. 
Collisional models have become a versatile framework for describing open quantum dynamics. 
Only recently have such models been formulated in a thermodynamically consistent form that ensures proper relaxation toward a Gibbs state when coupled to a single heat bath at fixed temperature~\cite{Jacob2021Apr, Tabanera2022Feb}. 

This approach reveals the crucial role of the energy-level spacing. 
In Refs.~\cite{Jacob2021Apr,Tabanera2022Feb}, we analyzed the effect of a single collision between a fixed system and a particle emerging from a thermal bath at equilibrium. 
A well-defined collision requires the incident particle to be localized in space, which, by the Heisenberg uncertainty principle, implies a finite momentum dispersion. 
The most general spatially localized states can be described as mixtures of wave packets, conveniently expressed through Wigner functions~\cite{Jacob2022Jun,Tabanera-Bravo2023May}. 
The effect of the collision depends sensitively on this momentum dispersion. 
We distinguish between {\em narrow packets}~\cite{Jacob2021Apr}, whose small momentum dispersion allows one to resolve the outgoing wave packets associated with distinct transitions of the system, and {\em broad packets}, whose large dispersion causes these outgoing packets to overlap. 
Such overlap leads to coherences between eigenstates of the Hamiltonian with nearly degenerate energies. 
In the limit of very large dispersion, the collision induces a purely unitary evolution of the system~\cite{Jacob2022Jun}, whereas collisions with narrow packets result in dissipative dynamics.

In this paper, we derive the evolution equation for extended systems locally bombarded by thermal particles. 
Specifically, we consider a chain of interacting qubits, where only the first qubit undergoes collisions with particles drawn from a thermal bath, and $\epsilon$ denotes the strength of the inter-qubit coupling. 
As expected, when the incident particles have negligible momentum dispersion $\sigma_p$, the resulting dynamical map remains global even as $\epsilon \to 0$. 
Conversely, if we take the limit $\epsilon \to 0$ while keeping $\sigma_p$ finite, the collisional map and the corresponding Lindblad equation become local. 
The noncommutativity of these two limits, $\sigma_p, \epsilon \to 0$, delineates the regimes in which the system’s evolution is dominated by either global or local contributions, depending on the energy-level spacing.


The paper is organized as follows. 
In Sec.~\ref{sec:cptp}, we provide a pedagogical overview of completely positive and trace-preserving (CPTP) maps and their connection to Lindblad equations. 
In Sec.~\ref{sec:lvsg}, we introduce precise definitions of local and global dynamical maps and generators, while Sec.~\ref{sec:therm} specifies the conditions under which these maps lead to thermalization. 
Section~\ref{sec:coll} analyzes the map arising from collisions with particles drawn from an equilibrium reservoir and shows that a global Lindblad equation emerges when the momentum variance of the incoming particles is small. 
In Sec.~\ref{sec:example}, we study in detail one-dimensional chains of qubits bombarded by particles with finite momentum variance and demonstrate that a local Lindblad equation is recovered in the limit of vanishing inter-qubit coupling. 
That section also assesses the validity of the different approximations through extensive numerical simulations. 
Finally, Sec.~\ref{sec:conclusions} summarizes the main conclusions of the work.

\section{CPTP maps and Lindblad equations}
\label{sec:cptp}

Consider a CPTP map and its Kraus representation:
\begin{equation}\label{kraus0}
    \mathbb{S}\rho=\sum_l M_l \rho M^\dagger_l,
\end{equation}
where $\sum_l M_l^\dagger M_l=\mathbb{I}$. The Kraus representation is not unique, but it provides an interesting interpretation of the map as a generalized non-selective measurement of a magnitude whose possible outcomes are $l$. On a pure state $\ket{\psi}$, each value $l$ is obtained with a probability
\begin{equation}
    p_l=\braket{\psi|M_l^\dagger M_l|\psi}=|| M_l\ket{\psi}||^2.
\end{equation}
When the outcome of the measurement is $l$, the system collapses to a new pure state
\begin{equation}
    \ket{\psi'}_l=\frac{M_l\ket{\psi}}{\sqrt{p_l}}.
\end{equation}
The condition $\sum_l M^\dagger_lM_l=\mathbb{I}$ implies that the probabilities $p_l$ are normalized. Notice also that unitary evolution is a particular case of CPTP map with a single unitary Kraus operator.

There are at least two ways of obtaining a continuous-time Lindblad equation from the map $\mathbb{S}$. The usual one is to consider a rapid concatenation of a map that is close to the identity $\mathbb{I}$, i.e., a map with Kraus operators of the form:
\begin{align}
\tilde    M_0&=\mathbb{I}-\left[\frac{i}{\hbar} H+\frac{\Gamma}{2}\,\sum_{l\neq 0}L_l^\dagger L_l\right]\Delta t \label{m0}\\
  \tilde  M_l& =\sqrt{\Gamma\Delta t} L_l \qquad (l\neq 0)\label{ml}
\end{align}
where $L_l$ are arbitrary operators, $H$ is a self-adjoint operator, and $\Gamma$ is a constant related to the probability to obtain $l\neq 0$ per unit of time.
The concatenation of this map in the limit $\Delta t\to 0$, yields the Lindblad equation
\begin{equation}\label{lindblad1}
    \dot\rho=-\frac{i}{\hbar}[H,\rho]+ \Gamma\sum_{l\neq 0}\left[L_l \rho L^\dagger_l- \frac{1}{2}\{\rho,L^\dagger_lL_l\}\right]
\end{equation}
where $\{A,B\}=AB+BA$ is the anticommutator.

An alternative way of obtaining a Lindblad equation from the CPTP map $\mathbb{S}$ defined in \eqref{kraus0} is assuming that the map acts at random Poissonian times occurring at a rate $\Gamma$, and the evolution between these times is unitary under Hamiltonian $H$ \cite{Strasberg2017,Ciccarello2022period,Tabanera-Bravo2023May}:
\begin{align}
    \dot\rho &=-\frac{i}{\hbar}[H,\rho]+\Gamma (\mathbb{S}-\mathbb{I})\rho \nonumber \\
   &= -\frac{i}{\hbar}[H,\rho]+\Gamma
    \sum_l\left[M_l \rho M_l^\dagger- \frac{1}{2}\{\rho,M^\dagger_lM_l\}\right].
    \label{lindblad0}
\end{align}

Notice that the sum of anticommutators is just twice $\rho$, since $\sum_l M_l^\dagger M_l$ is the identity. Hence, \eqref{lindblad0} is a very peculiar Lindblad equation. The difference between \eqref{lindblad1} and \eqref{lindblad0} is the third term in the r.h.s. of \eqref{m0}.

The physical interpretation is clearer if 
we write the Kraus operators that yield the Lindblad equation \eqref{lindblad0} as in \eqref{m0} and \eqref{ml}:
\begin{align}
\tilde    M_0&=\mathbb{I}-\left[\frac{i}{\hbar} H+\frac{\Gamma}{2}\mathbb{I}\right]\Delta t \simeq \sqrt{1-\Gamma\Delta t}\left[\mathbb{I}-\frac{i}{\hbar}\,H\Delta t\right]\nonumber \\
  \tilde  M_l& =\sqrt{\Gamma\Delta t} M_l.
  \label{mspois}
\end{align}
These operators correspond, respectively, to a unitary evolution occurring during a time interval of duration $\Delta t$ with probability $p_0=1-\Gamma\Delta t$, and
a non-selective measurement with outcomes $l$, occurring with probability $\Gamma\Delta t ||M_l\ket{\psi}||^2$. This is equivalent to performing a non-selective measurement at Poissonian times, as mentioned above.

Notice the physical difference between the two Lindblad equations.
Eqs.~\eqref{m0}-\eqref{ml} can be interpreted as describing a continuous measurement with a {\em null} outcome $l=0$. A typical example is a detector that monitors transitions in the system by detecting emitted or absorbed photons, or by continuously measuring an ancilla. Outcomes with $l\neq 0$ correspond to the detection of specific transitions, while $l=0$ indicates their absence. 

During a time $\Delta t$, such transitions occur with probability of order $\Gamma\Delta t$; that is, they follow a Poisson process with rate $\Gamma$ \cite{breuer2007}.  The key difference between Eq.~\eqref{mspois} and Eqs.~\eqref{m0}-\eqref{ml} is that, in the former, the system evolves unitarily between measurements, whereas in the latter, the evolution between transitions is continuously affected by the measurements, even when the outcome is $l=0$.

As already mentioned, the only difference between \eqref{lindblad1} and \eqref{lindblad0} is the normalization condition $\sum_l M_l^\dagger M_l=\mathbb{I}$. Throughout the paper, we will use the notation $M_l$ for operators satisfying this condition and $L_l$ for general Lindblad operators. Recall that the former can be interpreted as Kraus operators of a CPTP map and as Lindblad operators in the equation \eqref{lindblad0} resulting from applying the map at random Poissonian times. For this reason, we also will refer to Eq.~\eqref{lindblad0} as a {\em Poissonian Lindblad equation}.

\section{Local and global maps}
\label{sec:lvsg}

In this section, we provide precise definitions of local and global maps. Our classification differs slightly from that introduced by Cattaneo {\em et al.} \cite{Cattaneo_2019}. Theirs is based on the dissipators obtained in weak coupling theory, whereas ours is based on the effect of maps on extended systems.

Consider a spatially extended system formed by subsystems 
$i=1,2,\dots,N$, with global Hamiltonian
$H$.
This Hamiltonian $H$ is an operator defined on the global Hilbert space ${\cal H}={\cal H}_1\otimes \dots\otimes {\cal H}_N$, 
with eigenstates $\ket{j}$ and eigenvalues $\epsilon_j$:
\begin{equation}
    H\ket{j}=\epsilon_j\ket{j}.
\end{equation}
This paper restricts the discussion to Hamiltonians with non-degenerate energy levels. We caution, however, that generalization to degenerate levels is not trivial.

\subsection{Local maps}
\label{localsing}

We define \textit{generalized local} operators $L$ (which could be Kraus or Lindbladian operators) as those that affect only the first subsystem, i.e., those that satisfy
\begin{align}\label{condlocal2}
    \braket{s_1',\dots,s_N'|L|s_1,\dots s_N}=&\lambda(\{s_i'\},\{s_i\})\braket{s_1'|L^{(1)}|s_1}\nonumber \\ &\times\delta_{s_2',s_2}\dots \delta_{s_N',s_N}.
\end{align}
Here $\{\ket{s_1,\dots,s_N}\equiv \ket{s_1}\otimes\dots\otimes \ket{s_N}\}$ is a local basis, where for each $l$, $\{\ket{s_l}\}$ is a basis for ${\cal H}_l$, 
$L^{(1)}$ is a local operator defined on the Hilbert space ${\cal H}_1$ of the first subsystem, 
and $\lambda(\{s_i'\},\{s_i\})$ is a complex factor that can depend on the global properties of the two states involved. 
If this factor is constant, $\lambda(\{s_i'\},\{s_i\})=1$, then the operator $L$ is \textit{strictly local}:
\begin{equation}\label{condlocal}
    L=L^{(1)}\otimes \mathbb{I}\otimes \dots \otimes \mathbb{I}.
\end{equation}
By including the factor $\lambda(\{s_i'\},\{s_i\})$, we consider operations whose effect on the system is local, i.e., it only affects the first subsystem,  but this effect can depend on global properties. Notice, however, that the definition of local maps must be expressed in the local basis, as in Eq.~\eqref{condlocal2}. In Appendix \ref{sec:proof}, we show how this definition can be expressed in terms of the eigenbasis of $H$ under some conditions.
The jump operators in the semi-local Lindblad equations introduced in \cite{potts_thermodynamically_2021} are particular cases of \eqref{condlocal2}.

Strictly local maps and local Lindblad equations appear in a repeated interaction scenario where the system interacts with the environment for a short time $\tau$ \cite{breuer2007,Barra2015Oct}. 
Consider the total Hamiltonian
 \begin{equation}\label{ham0}
     H_{\rm tot}=H\otimes\mathbb{I}_E+\mathbb{I}\otimes H_E+\nu V\otimes {B}
 \end{equation}
 where $H_E$ is the Hamiltonian of the environment, and $\nu$ is the intensity of the interaction. $V$ and ${B}$ are self-adjoint operators defined on the Hilbert space of the system ${\cal H}$ and the environment ${\cal H}_E$, respectively. The effect of the interaction on the system reads
 \begin{equation}
     \rho'={\rm Tr}_E\left[e^{-i\tau H_{\rm tot}/\hbar}(\rho\otimes \rho_E)\, e^{i\tau H_{\rm tot}/\hbar}\right]\label{rhop}
 \end{equation}
 where $\rho_E$ is the same state for all interactions. We further assume that ${\rm Tr}_E[\rho_E {B}]=0$.
 
For short interactions $\tau\to 0$ and strong  coupling $\nu\to\infty$ scaling as $g\equiv\nu\sqrt{\tau}$, the unitary evolution can be approximated by \cite{karevski2009,Barra2015Oct,DeChiara2018Nov}
\begin{align}
     e^{-i\tau H_{\rm tot}/\hbar} \simeq &\mathbb{I}-\frac{i\tau}{\hbar}(H\otimes\mathbb{I}_E+\mathbb{I}\otimes H_E)-\frac{ig\sqrt{\tau}}{\hbar}V\otimes {B} \nonumber \\
     &-\frac{g^2\tau}{2\hbar^2}(V\otimes {B})^2\label{rhopexp}
\end{align}   
Introducing \eqref{rhopexp} in \eqref{rhop}, the terms of order $\sqrt{\tau}$ vanish because ${\rm Tr}_E[\rho_E {B}]=0$, and we obtain
 \begin{align}
 \rho' \simeq & \rho- \frac{i\tau}{\hbar}[H,\rho]+\frac{g^2\tau}{\hbar^2}\,{\rm Tr}_E\left[ V\otimes {B} (\rho\otimes \rho_E )V\otimes {B}\right]\nonumber \\
 &-\frac{g^2\tau}{2\hbar^2}\,{\rm Tr}_E\left[ \{ (V\otimes {B})^2,(\rho\otimes \rho_E )\}\right].\label{lindshort}
 \end{align}
 In the limit $\tau\to 0$, the resulting equation is \eqref{lindblad1} with $L_l=V$ (and no summation over $l$) and 
 \begin{equation}
    \Gamma=\frac{g^2}{\hbar^2}{\rm Tr}_E[{B}^2\rho_E].
 \end{equation}
Hence, if $V$ is a local operator, the repeated interaction scheme yields a \textit{local Lindblad equation} in the limit $\tau\to 0$. The previous argument can be extended to interaction terms of the form $\sum_i V_i\otimes B_{i}$, with $V_i$ local, which yield \textit{strictly local Lindblad equations} with several Lindblad operators \cite{Barra2015Oct,DeChiara2018Nov}.

However, as mentioned in the introduction, these Lindblad equations typically do not induce thermalization due to the work done by the external agent that switches the interaction on and off. 

We also note that local Lindblad equations have been derived within weak coupling theory under different assumptions, e.g. high temperature \cite{Esposito2005-kf} or high bias limit \cite{Harbola2006-cf}, where thermalization becomes trivial or irrelevant.

\subsection{Global maps and coherences}

On the other hand, when referring to \textit{global Lindblad equations}, we typically denote those derived in the framework of weak coupling theory and based, equivalently, on the Davis procedure \cite{Davies1974-ho, davis1976, SpohnLebowitz1978, Spohn1980-vb} or the secular approximation \cite{CohenTBook, breuer2007, rivas2012, Strasberg2022-lu}.
One of the main features of these equations is the decay of coherences in the eigenbasis of the Hamiltonian $H$. The effect of CPTP maps on coherences has been extensively studied in the context of resource theory \cite{chitambar2016,streslov2017}. 
To study this effect, it is convenient to express the map in the eigenbasis of the system Hamiltonian $\{\ket{j}\}$. If $\rho_{jk}\equiv \braket{j|\rho|k}$, the components of the map, $\mathbb{S}_{j'k'}^{jk}$, are given by
 \begin{equation} \label{stensor}   
 \braket{j'|\mathbb{S}\rho|k'}=
 \sum_{jk}\mathbb{S}_{j'k'}^{jk}\rho_{jk}.
 \end{equation}
In terms of the Kraus representation \eqref{kraus0}, we find
\begin{equation}\label{stensor2}
   \mathbb{S}_{j'k'}^{jk}=\sum_l \braket{j'|M_l|j} \braket{k'|M_l|k}^*.
\end{equation}

\textit{Incoherent operations} (IO) are maps that admit a Kraus representation where the individual operators do not create coherences, i.e., if $\rho$ is diagonal in the eigenbasis of $H$, then $M_l\rho M_l^\dagger$ is also diagonal in that basis (recall that we restrict the discussion to non-degenerate Hamiltonians and, therefore, the eigenbasis of $H$ is unique). Consequently,
\begin{equation}
    \braket{j'|M_l|j} \braket{k'|M_l|j}^*=0\qquad \mbox{if $j'\neq k'$.}
\end{equation}
This is equivalent to imposing that $M_l$ maps eigenstates into eigenstates of the Hamiltonian $H$. Then, the Kraus operators are of the form
\begin{equation}\label{io}
    M_l=\sum_j m^l_j\ket{k_l(j)}\bra{j} \;,
\end{equation}
where $k_l(j)$ is a function from the index set of the Hamiltonian eigenstates to itself, and $m^l_j$ are complex coefficients \cite{winter_operational_2016}.
These CPTP maps, and the corresponding Lindblad equations they generate, are commonly referred to as global, since the associated jump operators induce transitions between eigenstates of the global Hamiltonian $ H$. In contrast to local maps, these global jumps can affect the system as a whole.

As an illustration, consider the qubit chain discussed in Sec.~\ref{sec:example}. 
The ground state $\ket{0}$ is local, whereas the first excited states are superpositions of states with a single excitation that can be anywhere in the chain. The operator $M_l$ maps $\ket{0}$ onto $\ket{k_l(0)}$, for which there is a non-zero probability of finding the excitation at an arbitrary distance from the first qubit. Consequently, although the system is in contact with the reservoir through the first qubit, the excitation is instantaneously spread over the whole chain, irrespective of its length.

\textit{Strictly incoherent operations} (SIO) are such that both $M_l$ and $M^\dagger_l$ are IO, i.e., when the function $k_l(j)$ in \eqref{io} is one-to-one \cite{winter_operational_2016}.

Another interesting class of maps are the so-called \textit{translational invariant operations} (TIO) \cite{streslov2017}, which obey:
\begin{equation}
    e^{-iHt}\mathbb{S}\rho e^{iHt}=\mathbb{S}\left[e^{-iHt}\rho e^{iHt}\right].
\end{equation}
In terms of the components of a map in the eigenbasis of the Hamiltonian $H$, TIO implies
\begin{equation}
  \mathbb{S}^{jk}_{j'k'}\neq 0 \quad\Longrightarrow\quad \Delta_{kj}=\Delta_{k'j'}  \;,
\end{equation}
where $\Delta_{kj}\equiv\epsilon_k-\epsilon_j$. In this case, the Kraus operators can be labeled by a subindex $s$ and an energy $\omega$, which runs over all possible Bohr energies $\Delta_{jk}$ in the system. These Kraus operators are of the form
\begin{equation}\label{condglobal}
M_{s,\omega}=\sum_{j'j}\gamma^{(s)}_{j'j}\ket{j'}\delta_{\Delta_{j'j},\omega}\bra{j},
\end{equation}
where $\delta$ is the Kronecker delta. Comparing \eqref{condglobal} and \eqref{io}, we conclude that TIO maps are IO and, if the Hamiltonian is non-degenerate, they are SIO as well \cite{streslov2017}. The operators $M_{s,\omega}$  induce jumps between the eigenstates of $H$ with a given energy difference $\omega$, i.e., they can be considered ladder operators satisfying \cite{breuer2007}:
\begin{equation}
    [H,M_{s,\omega}]=\omega M_{s,\omega}.
\end{equation}

The Lindbladian operators derived from weak coupling theory are of the form given by Eq.~\eqref{condglobal} \cite{breuer2007,rivas2012}.
Notice that, if the maps are contractive and the steady state is unique, TIO implies that coherences decay.

\subsection{Thermal maps}

The so-called \textit{thermal operations} are another interesting family of maps that can appear in repeated interaction scenarios. They result from the interaction of a system with a thermal environment under the assumption that the total energy is conserved. More precisely, consider a system and its environment with Hamiltonian $H_0=H\otimes \mathbb{I}_E+ \mathbb{I}\otimes H_E$. A \textit{thermal map} is given by \cite{streslov2017}
\begin{equation}\label{thermalmap}
    \mathbb{S}\rho={\rm Tr}_E [U\rho\otimes \rho^{(E)}_{\rm therm} U^\dagger] \;,
\end{equation}
where $U$ is a unitary operator that commutes with $H_0$, and
\begin{equation}
    \rho^{(E)}_{\rm therm}=\frac{e^{-\beta H_E}}{Z_E}
\end{equation}
is the thermal state of the environment, with 
$\beta$ being the inverse temperature and $Z_E={\rm Tr}\left[e^{-\beta H_E}\right]$ the partition function. It is straightforward to show that the map is TIO and that the thermal state of the system $\rho_{\rm therm}\equiv e^{-\beta H}/Z$ is
invariant under thermal maps, i.e., $\mathbb{S}\,\rho_{\rm therm}=\rho_{\rm therm}$. Moreover, if this invariant state is unique,  any initial condition relaxes to the thermal state.

Notice that the repeated scheme given by Eqs.~\eqref{ham0} and \eqref{rhop} would yield a thermal operation if $V\otimes B$ commutes with $H_0$. In this case, the operator $V$ induces jumps between eigenstates of $H$, which are globally entangled. This is a particular situation in which the effect of the interaction with the environment spreads throughout the system trivially and is not suitable for studying local interactions with the environment.

All the previous characterizations of operators 
and all the properties studied in this section apply to both Lindblad operators $L_l$ and Kraus operators $M_l$.

\section{Thermalization}
\label{sec:therm}

We expect that any system in contact with a single thermal bath at temperature $T$ will eventually thermalize at the same temperature. Otherwise, one could build a perpetuum mobile of the second kind \cite{Ehrich2020}. 

The Lindblad equations \eqref{lindblad1} and \eqref{lindblad0} induce thermalization if they bring any initial state to a unique stationary state given by the Gibbs state $\rho_{\rm therm}$.
Thermalization thus implies {\em i)} the decay of coherences in the Hamiltonian eigenbasis, {\em ii)} the evolution of populations towards the values given by the Boltzmann factor.
We have already seen that the first condition is fulfilled by IO contractive maps, and now we discuss the second one.


Let us consider the Lindblad equation \eqref{lindblad1} in the eigenbasis of the Hamiltonian when $\rho$ is diagonal in this basis:
\begin{align}
    \dot \rho_{jk}&=\Gamma\sum_{l}\sum_{j'}\Big[
    \braket{j|L_l|j'}\braket{k|L_l|j'}^*\rho_{j'j'}\nonumber \\
    &-\braket{j'|L_l|j}^*\braket{j'|L_l|k}
    \frac{\rho_{jj}+\rho_{kk}}{2}
    \Big].
\end{align}
If the Lindbladian operators $L_l$ are IO, the dynamics is not producing any coherences ($\dot \rho_{jk}=0$ when $j\neq k$) and the stationary state is therefore incoherent. 
For $j=k$, the equation reads:
\begin{equation}\nonumber
  \dot \rho_{jj}=\Gamma\sum_{l}\sum_{j'}\left[
    |\braket{j|L_l|j'}|^2\rho_{j'j'}-|\braket{j'|L_l|j}|^2
    \rho_{jj}
    \right], 
\end{equation}
which is a master equation for the populations $\rho_{jj}(t)$ with transition rates:
\begin{equation}
    \Gamma_{j\to j'}=\Gamma \sum_l|\braket{j'|L_l|j}|^2.
\end{equation}
If these rates obey detailed balance
\begin{equation}
    \frac{\Gamma_{j\to j'}}{\Gamma_{j'\to j}}=\frac{\sum_l|\braket{j'|L_l|j}|^2}{\sum_l|\braket{j|L_l|j'}|^2}=e^{-\beta\Delta_{j'j}},
\end{equation}
then thermalization is warranted.

Regarding equation \eqref{lindblad0}, which we recall is identical to \eqref{lindblad1} except for the normalization condition $\sum_l M^\dagger_lM_l=\mathbb{I}$, exact thermalization occurs if the operators $M_l$ are IO, contractive, and the detailed balance condition is fulfilled. In this case, detailed balance can be written as:
\begin{equation}\label{dbs}
    \frac{\mathbb{S}_{j'j'}^{jj}}{\mathbb{S}_{jj}^{j'j'}}=\frac{\sum_l|\braket{j'|M_l|j}|^2}{\sum_l|\braket{j|M_l|j'}|^2}=e^{-\beta\Delta_{j'j}}.
\end{equation}

The above conditions (contractive IO and detailed balance) imply exact thermalization. However, IO's are global maps, 
preventing the possibility of a crossover between local and global Lindblad equations.

On the other hand, we have recently shown 
in Ref.~\cite{Tabanera-Bravo2023May} that thermalization can be approximately reached even when the Kraus operators $M_l$ are not IO. Thermalization occurs if the map acts on random Poissonian times, as in the scenario leading to the Lindblad equation \eqref{lindblad0},  with a low rate $\Gamma$.  In that case, the only relevant condition is detailed balance \eqref{dbs}. The proof in \cite{Tabanera-Bravo2023May}, which is a perturbative analysis of the Lindblad equation \eqref{lindblad0}, can be straightforwardly extended to generic Lindblad equations \eqref{lindblad1}. We start by expanding the stationary density matrix as
\begin{equation}
    \rho=\rho^{(0)}+\Gamma\rho^{(1)}+\dots
\end{equation}
and inserting the expression in the Lindblad equation \eqref{lindblad1}. If $\rho$ is stationary, then $\dot \rho=0$, and collecting  the terms of zeroth and first order in $\Gamma$, we obtain
\begin{align}
   0 &= -\frac{i}{\hbar}[H,\rho^{(0)}]  \\
     0 &= -\frac{i}{\hbar}[H,\rho^{(1)}]+\sum_l\left[ L_l\rho^{(0)}L^\dagger_l-\frac{1}{2}\{ \rho^{(0)},L_l^\dagger L_l\}\right].\label{lindpert}
\end{align}
The first equation implies that $\rho^{(0)}$ is diagonal in the eigenbasis of the Hamiltonian $H$. The second equation leads to
\begin{equation}
  0= \sum_{j'}\sum_{l}\left[
    |\braket{j|L_l|j'}|^2\rho^{(0)}_{j'j'}-|\braket{j'|L_l|j}|^2
    \rho^{(0)}_{jj}
    \right], 
\end{equation}
where $\rho_{jk}^{(0)}=\braket{j|\rho^{(0)}|k})$. The solution of this equation is the thermal state, provided that the detailed balance condition \eqref{dbs} holds.

The first correction $\rho^{(1)}$ to the thermal state can be calculated in the perturbative analysis. We limit ourselves to the calculation of the off-diagonal terms in the case of Poissonian Lindblad equations. Multiplying Eq.~\eqref{lindpert} by $\bra{j}$ and $\ket{k}$, with $j\neq k$, we get
\cite{Tabanera-Bravo2023May}
\begin{align}
\rho_{jk}^{(1)}&=-\frac{i\hbar}{\Delta_{jk}}\sum_l\sum_{j'}\rho_{j'j'}^{(0)}\braket{j|M_l|j'}\braket{j'|M_l^\dagger|k}\nonumber \\
&=-\frac{i\hbar}{\Delta_{jk}}\sum_{j'}\mathbb{S}^{j'j'}_{jk}\rho_{j'j'}^{(0)}\label{pert1}
\end{align}
Here, we have replaced the Lindblad operators $L_l$ by Kraus operators $M_l$ satisfying $\sum_l M^\dagger_lM_l=\mathbb{I}$.

To summarize, thermalization of open quantum systems involves two aspects: {\em i)} decoherence in the eigenbasis of the Hamiltonian $H$, $\rho_{jk}\propto \delta_{jk}$ and {\em ii)} thermalization of populations, i.e., $\rho_{jj}=e^{-\beta \epsilon_j}/Z$. SIO Lindblad operators, or IO Kraus operators, exactly warrant the first condition and yield global Lindblad equations. On the other hand, an approximate way to destroy coherence is for the dissipator in the Lindblad equation \eqref{lindblad1} or \eqref{lindblad0} to be small, $\Gamma\to 0$, as our previous perturbative analysis shows. The accuracy of this perturbative solution has been studied in detail in Ref.~\cite{Tabanera-Bravo2023May}. Finally, detailed balance warrants the thermalization of populations.

We conclude that, to obtain a thermodynamically consistent and non-global evolution equation for extended systems, we should consider weak dissipators ($\Gamma$ small) that obey detailed balance \eqref{dbs}. In the following sections, we show that a repeated-interaction scheme, in which a system is bombarded by particles from a thermal bath, leads to this type of evolution, provided that the bombardment occurs at random Poissonian times with a low rate $\Gamma$.

\vspace{0.4cm}

\section{Collisional reservoirs}
\label{sec:coll}

\subsection{The collisional map}

In \cite{Jacob2021Apr, Tabanera2022Feb}, we analyzed the interaction between a system with Hamiltonian $H$ and particles coming from a reservoir. These particles, also called {\em units}, move in one dimension and can have internal degrees of freedom with a Hamiltonian $H_U$ \cite{Tabanera2022Feb}. The scenario is described by the total Hamiltonian
\begin{equation}\label{htot}
    H_{\rm tot}=\frac{\hat p^2}{2m}+H_{U}+H+ \chi_L(\hat x)\,\nu V\otimes {B}\end{equation}
where $\hat p$ and $\hat x$ are the momentum and position operators of the 
incident particle, respectively. The interaction  term between the unit and the system is proportional to $\chi_L(x)$, which is the indicator 
function of the interval
$[-L/2,L/2]$: 
$\chi_L(x)=1$ if $x\in [L/2,-L/2]$ 
and zero otherwise, implying that the interaction occurs only in the interval $[-L/2,L/2]$.  Outside this scattering 
region, the free Hamiltonian $H_0 = H_U + H$ rules the evolution of the internal degrees of freedom of the system and the unit. Within the scattering region, the Hamiltonian affecting the internal degrees of freedom is $H_0+\nu V\otimes {B}$, where $V$ is an operator defined on the Hilbert space of the system and $B$ is an operator defined on the Hilbert space of the internal degrees of freedom of the incident particle.


In reference \cite{Jacob2021Apr}, we calculated the collisional map using quantum scattering theory in one dimension and obtained the following expression for the scattering map in the eigenbasis of the Hamiltonian $H$:
\begin{widetext}
\begin{equation}\label{sbig}
    \mathbb{S}_{j'k'}^{jk} =  \sum_{\alpha=\pm} \int dp \, dp'' \rho_{\rm U}(p,p'')\frac{\sqrt{pp''}}{m} \delta(E_p-E_{p''} -\Delta_{j'j} +\Delta_{k'k})s^{(\alpha  )}_{j'j}(E_{p} + \epsilon_j)\left[s^{(\alpha )}_{k'k}(E_{p''} + \epsilon_k)\right]^* \;
\end{equation}
where $\rho_{\rm U}(p,p'')$ is the density matrix of the incident 
particle in the momentum representation, $m$ is the mass of the 
particle, and $E_p\equiv p^2/(2m)$. $s^{(\alpha)}_{j'j}
(E)$ is the scattering amplitude corresponding to a transition 
$j\to j'$ in the target system as a consequence of a collision where 
the total energy is $E$, the incident particle has positive momentum before the collision, and the sign of the momentum after the collision is $\alpha=\pm$, i.e., $\alpha=+$ corresponds to transmitted particles and $\alpha=-$ to reflected ones. These amplitudes depend on the interaction term $\nu V\otimes {B}$. Finally, the integrals run over positive momenta $p$ and $p''$ verifying
$E_p\geq \Delta_{j'j}$ and $E_{p''}\geq \Delta_{k'k}$.
Eq.~\eqref{sbig} can be further simplified by integrating over $p''$. The result is \cite{Jacob2021Apr}
\begin{equation}\label{sbig2}
    \mathbb{S}_{j'k'}^{jk} =  \sum_{\alpha=\pm} \int_{p_{\rm inf}}^\infty dp \, \rho_{\rm U}(p,\pi(p))\sqrt{\frac{p}{\pi(p)}}\, s^{(\alpha  )}_{j'j}(E_{p} + \epsilon_j)\left[s^{(\alpha)}_{k'k}(E_{p}-\Delta_{j'j} + \epsilon_{k'})\right]^* \;
\end{equation}
with $\pi(p)=\sqrt{p^2-2m(\Delta_{j'j}-\Delta_{k'k})}$ and $p_{\rm inf}$ verifies $p_{\inf}^2=2m\,\max\{0,\Delta_{j'j},\Delta_{j'j}-\Delta_{k'k}\}$. Notice that the collision tensor must obey the relationship
\begin{equation}\label{congS}
    \mathbb{S}^{jk}_{j'k'}=  \left[\mathbb{S}^{kj}_{k'j'}\right]^*
\end{equation}
This relationship is apparent in Eq.~\eqref{sbig}. In Eq.~\eqref{sbig2}, one has to make the change of variable $p'=\pi(p)$ to prove \eqref{congS}.
The collisional map given by \eqref{sbig} or \eqref{sbig2} admits a Kraus representation. The corresponding operators are derived in Appendix \ref{app:kraus}. They do not fall into either of the two categories: local or global. 

\vspace{0.2cm}


\subsection{Detailed balance}
\label{sec:detbal}

For the collisional tensor \eqref{sbig2} to satisfy detailed balance, the collision and the incident units must fulfill two conditions. First, 
the scattering matrix must be symmetric $s_{j'j}^{(\alpha)}(E)=s_{jj'}^{(\alpha)}(E)$, a condition equivalent to micro-reversibility, i.e., the commutation between the total Hamiltonian $H$ and an anti-unitary time-reversal operator \cite{Jacob2021Apr,Tabanera2022Feb}. Second, the internal state of the units must be thermal and their momentum must be distributed according to the effusion distribution $\mu_{\rm eff}(p)$:
\begin{equation}\label{meff}
    \rho_{\rm U}(p,p)=\mu_{\rm eff}(p)\equiv \frac{\beta p}{m} e^{-\beta p^2/(2m)}.
\end{equation} 
One can use Wigner functions to build these states, as explained in \cite{Jacob2022Jun} and \cite{Tabanera-Bravo2023May}. Assuming that the Wigner function factorizes and that the probability distributions for the momentum and the position are Gaussian, one obtains \cite{Tabanera-Bravo2023May}:
\begin{equation}\label{rumeff}
    \rho_{\rm U}(p,p')=\mu_{\rm eff}\left(\frac{p+p'}{2}\right)e^{-\frac{(p-p')^2}{2\sigma_p^2}},
\end{equation} 
where $\sigma_p$ can be interpreted as an effective standard deviation of the momentum in the pure states that conform $\rho_{\rm U}$.

To check that \eqref{meff} induces the detailed balance condition, we write the transition probabilities particularizing Eq.~\eqref{sbig2} for $k=j$ and $k'=j'$ and $\rho_{\rm U}(p,p)=\mu_{\rm eff}(p)$:
 \begin{equation}
    \mathbb{S}^{jj}_{j'j'}=\frac{\beta}{m}
    \sum_{\alpha=\pm}\int_{p^{\rm inf}_{j'j}}^\infty dp 
    \left|s^{(\alpha)}_{j'j}\left(\frac{p^2}{2m}+\epsilon_j\right)\right|^2
    \,p\,
    e^{-\beta p^2/(2m)}
\end{equation}
with $p^{\rm inf}_{j'j}=\sqrt{2m\Delta_{j'j}}$ if $\Delta_{j'j}\geq 0$ and $p^{\rm inf}_{j'j}=0$ otherwise. 

Using the symmetry of $s_{j'j}^{(\alpha)}$ and  changing  the integration variable to $p'=\sqrt{p^2-2m\Delta_{j'j}}$ with $dp'=p\,dp/p'$, we get
 \begin{align}
   \mathbb{S}^{jj}_{j'j'}&=\frac{\beta}{m}\sum_{\alpha=\pm}\int_{p^{\rm inf}_{jj'}}^\infty dp' \left|s^{(\alpha)}_{jj'}\left(\frac{p'^2}{2m}+\epsilon_{j'}\right)\right|^2
    \,p'\,
    e^{-\beta (p'^2/(2m)+\Delta_{j'j})}
    =\mathbb{S}^{j'j'}_{jj}e^{-\beta\Delta_{j'j}}
\end{align}
which is the detailed balance condition \eqref{dbs}.

\end{widetext}

\subsection{Narrow incident states in momentum representation yield global Lindblad equations}
\label{sec:narrow}


In Ref.~\cite{Jacob2021Apr}, we distinguished
between narrow and broad wave packets in momentum representation. The former are wave functions with a small momentum dispersion.  For mixed states, like \eqref{rumeff}, the distinction is based on the effective standard deviation $\sigma_p$.
Narrow incident states in momentum representations are those with $\sigma_p$ small compared to the change of momentum induced by all possible transitions in the system, i.e.:
\begin{equation}\label{rhounarrow}
    \rho_{\rm U}(p,\pi(p))=\begin{cases}
        0 & \mbox{if $\Delta_{j'j}\neq \Delta_{k'k}$.} \\
        \mu_{\rm eff}(p) & \mbox{if $\Delta_{j'j}= \Delta_{k'k}$}
    \end{cases}
\end{equation}
For narrow incident states, the collisional tensor \eqref{sbig2} becomes:
\begin{equation}\label{sbignarrow}
    \mathbb{S}_{j'k'}^{jk} =  \sum_{\alpha=\pm} \int_{p_{\rm inf}}^\infty dp \, \mu_{\rm eff}(p)\, s^{(\alpha  )}_{j'j}(E_{p} + \epsilon_j)\left[s^{(\alpha )}_{k'k}(E_{p} + \epsilon_{k})\right]^* 
\end{equation}
if $\Delta_{j'j}=\Delta_{k'k}$ and zero otherwise. In this expression, $p_{\rm inf}=\sqrt{2m\cdot\max\{0,\Delta_{j'j}\}}$. From \eqref{sbignarrow}, one can obtain a Kraus representation with operators $M_{\alpha,p,\omega}$, where $\omega$ runs over all possible Bohr energies and $p\in [p_{\rm inf},\infty)$. These operators read, in the eigenbasis of $H$,
\begin{equation}\label{mnarrow}
    \braket{j'|M_{\alpha,p,\omega}|j}=\sqrt{\mu_{\rm eff}(p)}\, s_{j'j}^{(\alpha)}(E_p+\epsilon_j)\,\delta_{\Delta_{j'j},\omega}
\end{equation}
which are TIO, as one infers by comparing \eqref{mnarrow} and \eqref{condglobal}.

Consequently, the resulting Lindblad equation is global and similar to the one obtained in Davies' weak coupling theory: it eliminates all coherences in the eigenbasis of $H$ and modifies the probability of each state according to a master equation that ensures detailed balance. In other words, it induces thermalization. 



\section{Case study: Locally bombarded qubit chains}
\label{sec:example}

\subsection{The model}

\begin{figure}
    \centering
    \vspace{0.35cm}
    \includegraphics[width=0.9\linewidth]{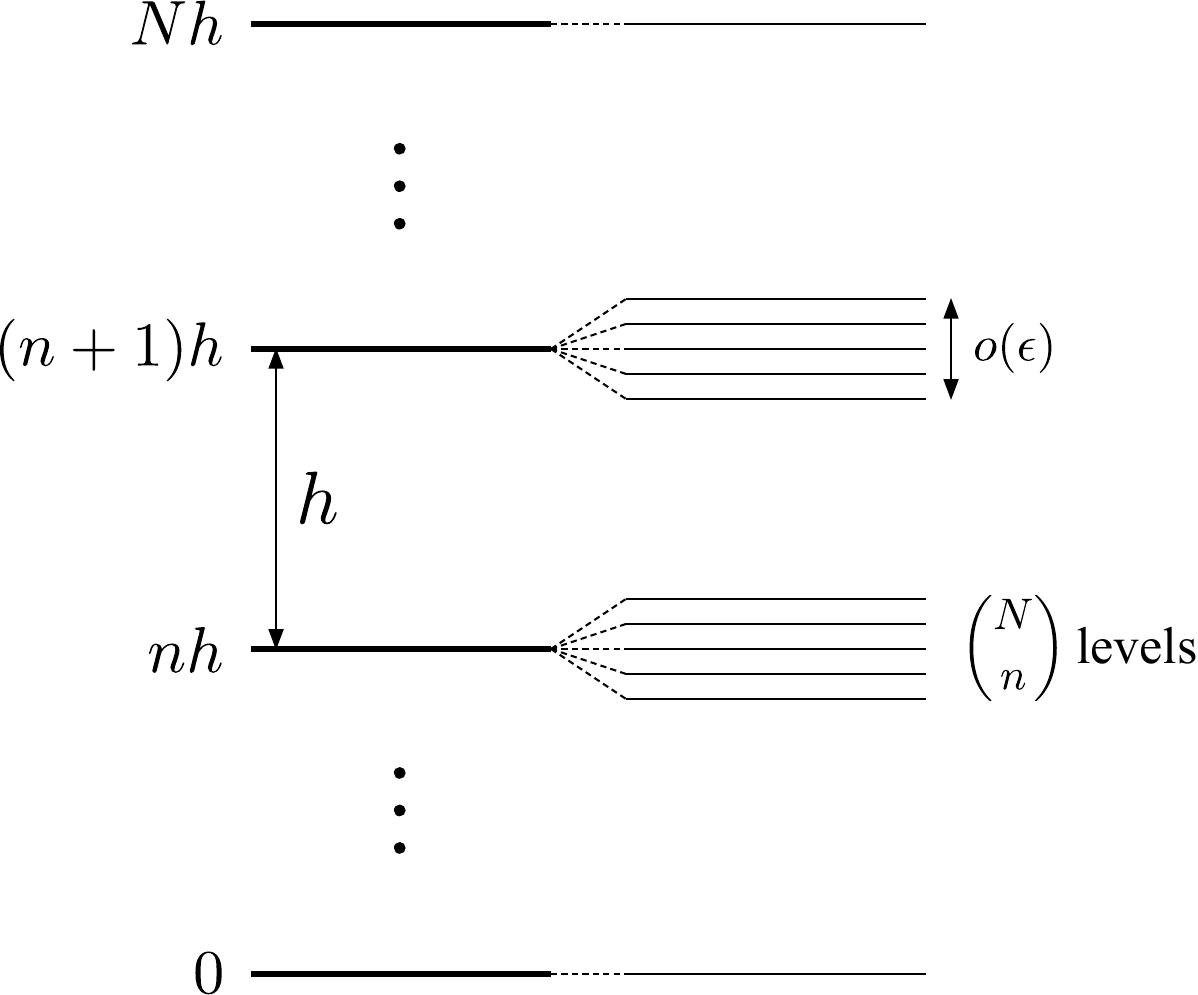}
    \caption{Spectrum of the spatially extended system with Hamiltonian $H$ given by Eq.~\eqref{hamext}. The energy levels of the local Hamiltonian $H_{\rm loc}$ are depicted as thick horizontal lines on the left. Their energy is $nh$, where $n$ is the number of excitations, and their degeneracy is $g_n\equiv\binom{N}{n}$. The perturbation $\epsilon H_{\rm int}$ for $\epsilon\ll h$ splits these levels into $g_n$ non-degenerate eigenstates of $H$. Hence, the energy levels of $H$ in Eq.~\eqref{hamext} are grouped in bands with $n$ excitations and $g_n$ non-degenerate levels.}
    \label{fig:spectrum}
\end{figure}

To study the locality of the scattering map \eqref{sbig2}, we consider a specific spatially extended system: a generic chain of qubits with a Hamiltonian
\begin{equation}\label{hamext}
    H=\sum_{i=1}^N h\frac{\sigma_z^{(i)}+\mathbb{I}^{(i)}}{2}+\epsilon  H_{\rm int}
    =H_{\rm loc}+\epsilon H_{\rm int} \;,
\end{equation}
where $\sigma^{(i)}_z$ and $\mathbb{I}^{(i)}$ are, respectively, the Pauli $z$-matrix and the identity operator for qubit $i$, and $H_{\rm int}$ is a non-local potential that accounts for the interaction between qubits. For $\epsilon\ll h$, a simple perturbative analysis tells us that the spectrum of $H$ is formed by bands labeled by an integer number $n=0,1,\dots,N$, corresponding to the number of excitations in the chain. Assuming that the interaction splits all the degenerate energy levels, there will be $\binom{N}{n}$ non-degenerate levels within the $n$-th band, whose energy has a dominant term $nh$ plus a shift of order $\epsilon$, as sketched in Fig.~\ref{fig:spectrum}.

If we bombard the first qubit, the scattering map $\mathbb{S}_{j'k'}^{jk}$ is given by \eqref{sbig2}. 
The scattering matrix $s_{j'j}^{(\alpha)}(E)$ will be local if the collision is short, as in the limit discussed in section \ref{sec:lvsg} or in the specific example analyzed below. 
However, the map is not local in general. On one hand, for incident particles with very small momentum dispersion obeying \eqref{rhounarrow}, the resulting Kraus operators \eqref{mnarrow} are global. These operators remain global even in the limit $\epsilon \to 0$. 
On the other hand, if the momentum dispersion $\sigma_p$ of the incident particles is finite, then \eqref{sbig} yields a local equation in the limit $\epsilon\to 0$, as expected. 
In other words, the limits $\sigma_p \to 0$ and $\epsilon \to 0$ do not commute. We explore this issue in the next subsections.

\vspace{0.4cm}

\subsection{Incident particles with small momentum dispersion}
   
As discussed previously, we should investigate incident particles with small but non-negligible momentum variance to study the crossover between local and global Lindblad equations.
A convenient initial state for the units is given by the following density matrix: 
\begin{equation}\label{rhoa2}
    \rho_{\rm U}(p,\pi(p))\simeq 
    \begin{cases}
    0 
    & \mbox{if $\Delta_{j'j}-\Delta_{k'k}=o(h)$}\\
    \mu_{\rm eff}\left(\frac{p+\pi(p)}{2}\right) 
    & \mbox{if $\Delta_{j'j}-\Delta_{k'k}=o(\epsilon)$}
    \end{cases}.
\end{equation}
Comparing this choice with \eqref{rhounarrow}, we see that \eqref{rhoa2}
describes incident particles that are narrow in momentum representation compared to the momentum change induced by a transition between bands, but broad compared to momentum changes induced by intra-band transitions.

This assumption is sufficient to destroy coherences between bands in the spectrum of $H$ since $\Delta_{j'j}-\Delta_{k'k}=\Delta_{j'k'}-\Delta_{jk}$ and the scattering operator given by \eqref{sbig2} yields
\begin{equation}
    \mathbb{S}^{jk}_{j'k'}\simeq 0 \quad \mbox{if $\Delta_{j'k'}-\Delta_{jk}$ is of order $h$.}
\end{equation}
Hence, if $j$ and $k$ are in the same band, $\Delta_{jk}$ is of order $\epsilon$, and $ \mathbb{S}^{jk}_{j'k'}=0$ for all $j',k'$ in different bands. In other words, intra-band coherences cannot create inter-band coherences.

 \begin{widetext}

We further assume that the scattering matrix is local, i.e., $s^{(\alpha)}_{j'j}(E)$ is different from zero only if $\ket{j}$ and $\ket{j'}$ are connected by a local operator. In this case, the scattering matrix can be written as
\begin{equation}
  s^{(\alpha)}_{j'j}(E) =\braket{j'|\left(\begin{array}{cc} a_0^{(\alpha)}(E) &  a_+^{(\alpha)}(E) \\ a_-^{(\alpha)}(E) & a_1^{(\alpha)}(E)\end{array}\right)\otimes \mathbb{I}\otimes\cdots\otimes \mathbb{I}|j}. 
\end{equation}
Notice that the entries of the matrix $a^{(\alpha)}_{0,1,\pm}(E)$ can depend on $j$ and $j'$. 
This matrix can be further decomposed into three terms:
\begin{equation}\label{matrixa2}
     s^{(\alpha)}_{j'j}(E) =\braket{j'|\left(\begin{array}{cc} a_0^{(\alpha)}(E) &  0 \\  & a_1^{(\alpha)}(E)\end{array}\right)\otimes \mathbb{I}\otimes\cdots\otimes \mathbb{I}|j}+a_+^{(\alpha)}(E)\braket{j'|\sigma_+^{(1)}|j}+a_-^{(\alpha)}(E)\braket{j'|\sigma_-^{(1)}|j}.
\end{equation}
Taking into account how the operators in each of these terms modify the first qubit, we find that
the first term is different from zero only when $\ket{j}$ and $\ket{j'}$ are in the same band, i.e., when $\Delta_{j'j}=o(\epsilon)$; the second term is different from zero only if $\Delta_{j'j'}=h+o(\epsilon)$; and the third one vanishes unless $\Delta_{j'j'}=-h+o(\epsilon)$.

Introducing this decomposition in Eq.~\eqref{sbig2} and taking into account \eqref{rhoa2}, the only non-zero elements of the collision tensor are the following. 
If $\Delta_{j'j}, \Delta_{k'k} = o(\epsilon)$:
\begin{align}
   \mathbb{S}^{jk}_{j'k'}=\sum_{\alpha=\pm}\int_{p_{\rm inf}}^\infty dp\,\sqrt{\frac{p}{\pi(p)}} &
\mu_{\rm eff}\left(\frac{p+\pi(p)}{2}\right)  
\braket{j'|\left(\begin{array}{cc} a_0^{(\alpha)}(E_p+\epsilon_j) & 0 \\ 0 & a_1^{(\alpha)}(E_p+\epsilon_j)\end{array}\right)\otimes \mathbb{I}\otimes\cdots\otimes \mathbb{I}|j}  \nonumber \\
\times &\braket{k'|\left(\begin{array}{cc} a_0^{(\alpha)}(E_p-\Delta_{j'j}+\epsilon_{k'}) & 0 \\ 0 & a_1^{(\alpha)}(E_p-\Delta_{j'j}+\epsilon_{k'})\end{array}\right)\otimes \mathbb{I}\otimes\cdots\otimes \mathbb{I}|k}\phantom{}^*
\label{sdecomp1}
\end{align}
where $p_{\rm inf}$ verifies $p_{\rm inf}^2=2m \max \{0,\Delta_{j'j},\Delta_{j'j}-\Delta_{k'k}\}$. These entries of the collision tensor govern the evolution within a band of the Hamiltonian spectrum.

If $\Delta_{j'j}=\Delta_{k'k}=\pm h +o(\epsilon)$, i.e., if the jumps involve  a change in the number of excitations, then:
\begin{equation}\label{sdecomp2}
 \mathbb{S}^{jk}_{j'k'}=\frac{1}{4}\,\sum_{\alpha=\pm}\int_{p_{\rm inf}}^\infty dp\,
\sqrt{\frac{p}{\pi(p)}}
\mu_{\rm eff}\left(\frac{p+\pi(p)}{2}\right) 
\left[ a_\pm^{(\alpha)}
(E_p+\epsilon_{j})\,
\braket{j'|\sigma_\pm^{(1)}|j}
\right]
\left[ a_\pm^{(\alpha)}
(E_p-\Delta_{j'j}+\epsilon_{k'})\,
\braket{k'|\sigma_\pm^{(1)}|k}
\right]^*   
\end{equation}
where   $\sigma_\pm^{(1)}\equiv \sigma^{(1)}_x\pm i\sigma^{(1)}_y$ are the ladder operators of the first spin. Here,
 $p_{\rm inf}^2=2m\Delta_{j'j}$ when $\Delta_{j'j}=h+o(\epsilon)$ and 
$p_{\rm inf}^2=2m\max\{0,\Delta_{j'j}-\Delta_{k'k}\}$ if $\Delta_{j'j}=-h+o(\epsilon)$. These entries only connect states that differ in the first qubit. However, they cannot be considered local even in the generalized sense \eqref{condlocal2}, because the amplitudes of these transitions depend on the eigen-energies $\epsilon_j$. Locality is only recovered in the limit $\epsilon\to 0$. Moreover, they obey detailed balance. Hence, as shown in Ref.~\cite{Tabanera-Bravo2023May}, Poissonian collisions at a low rate $\Gamma$ drive the system toward the thermal state at temperature $T$.


\subsection{The limit $\epsilon\to 0$.}
\label{sec:eps0}

In this limit, the scattering matrix and the scattering map are trivially local if the incident particles have a finite momentum dispersion. Nevertheless, it is interesting to calculate the scattering map in the eigenbasis of $H$, $\ket{j}$, and the corresponding local Kraus operators. The entries $\mathbb{S}^{jk}_{j'k'}$ of the collisional tensor depend now on $\epsilon^{(0)}_j$, the energy of state $\ket{j}$ for $\epsilon=0$, which is $h$ times the number of excitations of state $\ket{j}$. The limit $\epsilon \to 0$ in  Eqs.~\eqref{sdecomp1} and \eqref{sdecomp2} reads:  

\begin{itemize}
    \item If $\Delta_{j'j}, \Delta_{k'k} = o(\epsilon)$:
\begin{align}
\mathbb{S}^{jk}_{j'k'}=\sum_{\alpha=\pm}\int_0^\infty dp\,
\mu_{\rm eff}(p)  &
\braket{j'|
\left(
\begin{array}{cc} 
a_0^{(\alpha)}(E_p+\epsilon^{(0)}_j) & 0 \\ 0 & a_1^{(\alpha)}(E_p+\epsilon^{(0)}_j)
\end{array}
\right)
\otimes \mathbb{I}\otimes\cdots\otimes \mathbb{I}|j}  \nonumber \\ 
\times &
\braket{k'|\left(\begin{array}{cc} a_0^{(\alpha)}(E_p+\epsilon^{(0)}_{k}) & 0 \\ 0 & a_1^{(\alpha)}(E_p+\epsilon^{(0)}_{k})\end{array}\right)\otimes \mathbb{I}\otimes\cdots\otimes \mathbb{I}|k}\phantom{}^*
\label{sdecomp1b}
\end{align}
\item If $\Delta_{j'j},\Delta_{k'k}=h +o(\epsilon)$:
\begin{equation}\label{sdecomp2b}
 \mathbb{S}^{jk}_{j'k'}=\frac{1}{4}\,\sum_{\alpha=\pm}\int_{\sqrt{2mh}}^\infty dp\,
\mu_{\rm eff}(p) 
\left[ a_+^{(\alpha)}
(E_p+\epsilon^{(0)}_{j})\,
\braket{j'|\sigma_+^{(1)}|j}
\right]
\left[ a_+^{(\alpha)}
(E_p+\epsilon^{(0)}_{k})\,
\braket{k'|\sigma_+^{(1)}|k}
\right]^*   .
\end{equation}  
\item If $\Delta_{j'j},\Delta_{k'k}=-h +o(\epsilon)$:
\begin{equation}\label{sdecomp2bb}
 \mathbb{S}^{jk}_{j'k'}=\frac{1}{4}\,\sum_{\alpha=\pm}\int_0^\infty dp\,
\mu_{\rm eff}(p) 
\left[ a_-^{(\alpha)}
(E_p+\epsilon^{(0)}_{j})\,
\braket{j'|\sigma^{(1)}_{-}|j}
\right]
\left[ a_-^{(\alpha)}
(E_p+\epsilon^{(0)}_{k})\,
\braket{k'|\sigma_-^{(1)}|k}
\right]^*  .
\end{equation}

\end{itemize}

From these expressions for the collision tensor, we can obtain three families of  Kraus/Lindbladian operators:
\begin{align}
   \braket{j'| M_{\alpha,p,0}|j}&=\sqrt{\mu_{\rm eff}(p)}\,\braket{j'|\left(\begin{array}{cc} a_0^{(\alpha)}(E_p+\epsilon^{(0)}_j) & 0 \\ 0 & a_1^{(\alpha)}(E_p+\epsilon^{(0)}_j)\end{array}\right)\otimes \mathbb{I}\otimes\cdots\otimes \mathbb{I}|j}   & p\in &[0,\infty)\\
   \braket{j'| M_{\alpha,p,-}|j}&=\frac{\sqrt{\mu_{\rm eff}(p)}}{2}\,a_-^{(\alpha)}
(E_p+\epsilon^{(0)}_{j})\,
\braket{j'|\sigma^{(1)}_{-}|j}   & p \in &[0,\infty)\\
\braket{j'| M_{\alpha,p,+}|j}&=\frac{\sqrt{\mu_{\rm eff}(p)}}{2}\,a_+^{(\alpha)}
(E_p+\epsilon^{(0)}_{j})\,
\braket{j'|\sigma^{(1)}_{+}|j}  & p \in &[\sqrt{2mh},\infty).
\end{align}
These Kraus operators are now generalized local and can be expressed in the local basis $\{\ket{s_1,\dots,s_N}$ with $s_i=0,1$,  using the result \eqref{condlocal2app} derived in Appendix \ref{sec:proof}. The non-zero entries of the matrices of the Kraus operators are the following:
\begin{align}
   \braket{s_1,\dots,s_N| M_{\alpha,p,0}|s_1,\dots,s_N} & =\sqrt{\mu_{\rm eff}(p)}\, \left[a_1^{(\alpha)}(E_p+nh)\delta_{s_1,1}+a_0^{(\alpha)}(E_p+nh)\delta_{s_1,0}\right]    & p\in &[0,\infty)\\
   \braket{s_1-1,\dots,s_N| M_{\alpha,p,-}|s_1,\dots,s_N} & ={\sqrt{\mu_{\rm eff}(p)}}\,a_-^{(\alpha)}
(E_p+nh)\,
   & p \in &[0,\infty)\\
\braket{s_1+1,\dots,s_N| M_{\alpha,p,+}|s_1,\dots,s_N}
&={\sqrt{\mu_{\rm eff}(p)}}\,a_+^{(\alpha)}
(E_p+nh)\,
  & p \in &[\sqrt{2mh},\infty)
\end{align}
where $n$ is the number of excitations of state $\ket{s_1,\dots,s_N}$, i.e., $n=\sum_i s_i$. These operators induce jumps on the first qubit. The corresponding jump amplitudes obey detailed balance with respect to the local Hamiltonian $H_{\rm loc}$, as one can show by repeating the argument of subsection \ref{sec:detbal}.

\vspace{0.1cm}

\end{widetext}



We recall that the Lindblad equation resulting from the bombardment with particles with zero momentum dispersion is of the same form as Davies weak coupling theory and is global, even in the limit $\epsilon\to 0$ since it kills all coherences in the eigenbasis of $H$ and induces jumps between its eigenstates $\ket{j}$, which exhibit global entanglement. 
Instead, for weak bombardment rates, we have obtained an equation that satisfies detailed balance and thermalizes, but its nonlocal effects are of the order of the interaction between subsystems $\epsilon$. In the limit $\epsilon\to 0$, the collisional map is local and only affects the first qubit, which thermalizes at temperature $T$ with respect to the Hamiltonian $H_{\rm loc}$. 
The bombardment rate $\Gamma$ introduces a time scale which must be compared with the time scale associated with the intra-band transitions $\hbar/\epsilon$. Eq.~\eqref{pert1} indicates that the coherences in the eigenbasis of the Hamiltonian $H$ are of the order of $\Gamma\hbar/\epsilon$. This seems to compromise thermalization in the limit $\epsilon\to 0$. However, this is not the case since the sum in \eqref{pert1} is of order $\epsilon$. Indeed, if $\ket{j}$, $\ket{j'}$, and $\ket{k}$ are in the same band with energy $nh+o(\epsilon)$, one has for $\epsilon=0$,
\begin{equation}\label{sum}
    \sum_{j}\mathbb{S}_{jj}^{j'k'}\rho^{(0)}_{jj}=\frac{e^{-\beta\epsilon_{j'}}}{Z}\delta_{j'k'}.
\end{equation}
The proof is given in Appendix \ref{app:scat}.
Therefore, the dominant term in the coherences is of order $\Gamma$ and independent of $\epsilon$. This means that one can achieve approximate thermalization even for large systems, where the level spacing within bands is extremely small.

\subsection{An explicit example}

We now discuss an explicit example of a chain of qubits bombarded by particles with finite momentum dispersion. The total Hamiltonian $H_{\rm tot}$ is given by \eqref{htot}. The Hamiltonian of the system is
\begin{equation}\label{eq:chainmiltonian}
    H=\sum_{i=1}^N h\frac{\sigma_z^{(i)}+\mathbb{I}^{(i)}}{2}+\epsilon\sum_{i=1}^{N-1}\left[\sigma_+^{(i)}\otimes \sigma_-^{(i+1)}+{\rm c.c.}\right]
\end{equation}
where $\sigma_\alpha^{(i)}$ with $\alpha=x,y,z$ are the Pauli matrices corresponding to qubit $i$, and $\sigma_\pm^{(i)}=\sigma_x^{(i)}\pm i\sigma_y^{(i)}$ are ladder operators.

As discussed above, if $\epsilon\ll h$, the spectrum of $H$ consists of bands separated by an energy of the order of $h$. The eigenstates of $H$ generally involve global entanglement and have an energy spacing of the order of $\epsilon$ within each band (see Fig.~\ref{fig:spectrum}). 

For simplicity, we consider incident particles with no internal degrees of freedom, and the interaction with the first qubit in the chain is localized at the origin $x=0$:
\begin{equation}\label{eq:unitmiltonian}
   H_{\rm US}= \frac{g}{2}\left[\sigma^{(1)}_++\sigma^{(1)}_-\right]\otimes \delta(\hat x)=g\,\sigma_x\otimes \delta(\hat x).
\end{equation}
where $\delta(x)$ is the Dirac delta \cite{Jacob2021Apr}. This interaction potential results from \eqref{htot} in the limit $L\to 0$ and $\nu\to\infty$ with $L\nu=g$.

In Appendix \ref{app:scat}, we explicitly calculate the scattering map $\mathbb{S}^{jk}_{j'k'}$ for $\epsilon\to 0$. The map is local and obeys detailed balance, as expected. More interestingly, we prove that
\begin{equation}
    \sum_j \mathbb{S}^{jj}_{j'k'}\, e^{-\beta\epsilon^{(0)}_j}=
    e^{-\beta\epsilon^{(0)}_{j'}}\delta_{j'k'}.
\end{equation}

This equation implies that the first-order term \eqref{pert1} in our perturbative analysis of section \ref{sec:therm} is of order $\Gamma$ and not $\Gamma/\epsilon$, since the sum in \eqref{pert1} is of order $\epsilon$ for $j\neq k$.

Moreover, in our specific model, the interaction commutes with the local Hamiltonian, $[H_{\rm loc},H_{\rm int}]=0$. Hence, the state $e^{-\beta H_{\rm loc}}/Z$ commutes with the total Hamiltonian $H$, and is the stationary solution of the Lindblad equation when we set $\epsilon=0$ in the scattering map.

To summarize these results, we make explicit the dependency of the Hamiltonian $H(\epsilon)$ and the dissipator $\mathbb{S}(\epsilon)$ on the interaction strength between qubits $\epsilon$:

\begin{itemize}

\item The exact and unique stationary solution of the local equation:
\begin{equation}\label{locals}
   \dot\rho= -\frac{i}{\hbar}[H(\epsilon),\rho]+\Gamma\left(\mathbb{S}(0)-\mathbb{I}\right)\rho
\end{equation}
is, for all $\Gamma$:
\begin{equation}
    \rho_{\rm ss}=\frac{e^{-\beta H(0)}}{Z_0}
\end{equation}

\item The unique stationary solution of the exact Lindblad equation:
\begin{equation}\label{lindexact}
  \dot\rho= -\frac{i}{\hbar}[H(\epsilon),\rho]+\Gamma\left(\mathbb{S}(\epsilon)-\mathbb{I}\right)\rho
\end{equation}
is:
\begin{equation}
    \rho_{\rm ss}=\frac{e^{-\beta H(\epsilon)}}{Z_\epsilon} + o(\Gamma).
\end{equation}

\end{itemize}

In the first equation, the dissipator is local, but the system does not thermalize to the complete Hamiltonian. This is a consequence of the thermodynamic inconsistency of the local dissipator pointed out in a number of references \cite{DeChiara2018Nov}. On the other hand, the exact dissipator becomes local when $\epsilon\to 0$ and drives the system towards the thermal state for low bombardment rates $\Gamma$.

\begin{figure*}
    \centering
    \includegraphics[width=\linewidth]{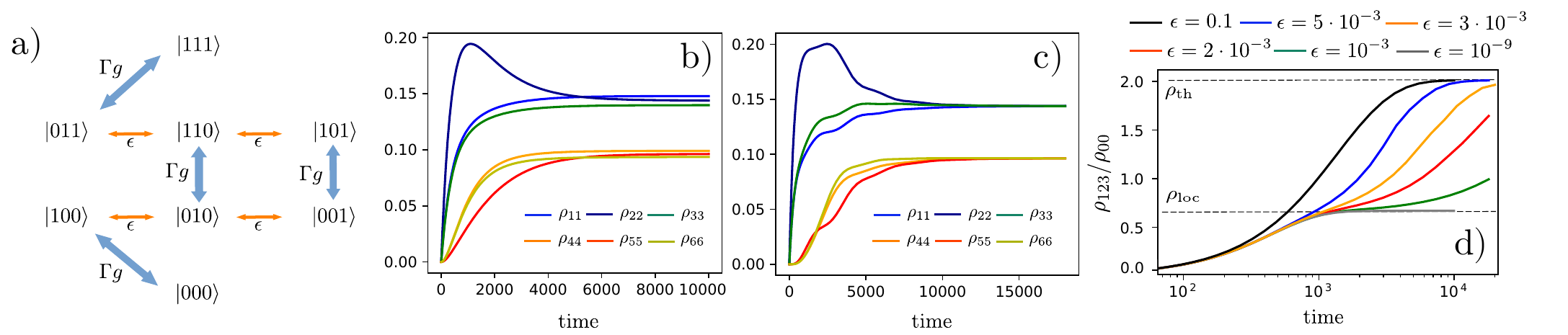} 
    \caption{Exact dynamics of the 3-qubit chain given by the Lindblad equation \eqref{lindblad0} with the exact 
    collisional map \eqref{sbig2}: 
    (a) Sketch of transitions between local states induced by {\em i)} the Hamiltonian Eq.\eqref{eq:chainmiltonian} (orange arrows), and {\em ii)} by collisions with the reservoir \eqref{eq:unitmiltonian} (blue arrows). 
    (b,c) Evolution of the eigensate populations $\rho_{ii}$ for $\epsilon = 0.1$ in (b) and $\epsilon = 5\times 10^{-5}$ in (c). 
    (d) Occupation of the first band, $\rho_{123} \equiv \rho_{11}+\rho_{22}+\rho_{33}$, relative to 
    the ground state, $\rho_{00}$, for different values of $\epsilon$. The two dashed gray lines give the value of the ratio for the global and local thermal states $\rho_{\rm therm}$, given by  Eq.~\eqref{stateglobal}, and  $\rho_{\rm loc}$, given by \eqref{statelocal}, respectively. Other parameters: $\beta = m = 0.1$, $g = 50$, $\Gamma = 1$, $\sigma_p = 0.5$ and $h = 4$.}
    \label{fig:termalizacion}
\end{figure*}

\subsection{Numerical results}

We study three Lindblad equations:
\begin{equation}\label{locals2}
    \dot\rho= -\frac{i}{\hbar}[H(\epsilon),\rho]+\Gamma\left(\mathbb{S}(\epsilon,\sigma_p)-\mathbb{I}\right)\rho
\end{equation}
resulting from bombardment at a rate $\Gamma$ and with the following versions of the scattering map $\mathbb{S}$: {\em i)} the 
exact map $\mathbb{S}(\epsilon,\sigma_p)$ given by \eqref{sbig2}; {\em ii)} the map for narrow packets, $\mathbb{S}(\epsilon,0)$, given by \eqref{sbignarrow} and yielding a global Lindblad equation; 
and {\em iii)} the {\em local} dissipator $\mathbb{S}(0,\sigma_p)$, given by  Eqs.~\eqref{sdecomp1b}--\eqref{sdecomp2bb} and  yielding the local Lindblad equation Eq.~\eqref{locals}. 

Notice that the {local} equation is \eqref{locals}, i.e., it is obtained by setting $\epsilon=0$ in the scattering map but keeping $\epsilon\neq 0$ in the Hamiltonian $H$. This is equivalent to neglecting terms of order $\Gamma\epsilon$ in the exact evolution 
equation. The resulting equation has a local dissipator, although the bare Hamiltonian $H$ is still non-local.


We have performed numerical simulations of a 3-qubit chain, $N=3$, for which the Hilbert space is given by linear combinations of local states $\ket{s_1\,s_2\,s_3}$, with $s_i=0,1$. The Hamiltonian \eqref{eq:chainmiltonian} commutes with the number of excitations $n\equiv s_1+s_2+s_3$ and its spectrum has the form depicted in Fig.~\ref{fig:spectrum}, with four bands corresponding, respectively, to $n=0,1,2$ and $3$.

The local states are affected both by collisions and by the unitary evolution between 
collisions, sketched in Fig.~\ref{fig:termalizacion}\,(a). The main effect of the 
former, ruled by the interaction Hamiltonian \eqref{eq:unitmiltonian} and the corresponding scattering map, is local flips in the first qubit, which result in 
transitions between bands. These transitions occur at a rate $\Gamma g$ and are depicted as 
blue arrows in Fig.~{\ref{fig:termalizacion}\,(a). On the other hand, the free evolution 
under Hamiltonian \eqref{eq:chainmiltonian} preserves the number of excitations, 
connecting the states within the bands $\left\{\ket{100},\ket{010},\ket{001} \right\}$ 
and $\left\{\ket{101},\ket{110},\ket{011} \right\}$. The intensity of this effect is given by  
the coupling constant between qubits, $\epsilon$, and is sketched as orange arrows in 
Fig.~\ref{fig:termalizacion}\,(a).

In Fig.~\ref{fig:termalizacion} (b-d), we plot the evolution of the density matrix elements $\rho_{ii} = \bra{i}\rho\ket{i}$ in the eigenbasis of the chain Hamiltonian $H$, Eq.~\eqref{eq:chainmiltonian}, i.e., the populations of the different eigenstates of $H$. These eigenstates are denoted as $\ket{i}$ for $i=0,\dots,7$, with $\ket{0}=\ket{000}$, $\ket{7}=\ket{111}$. The eigenstates $\ket{i}$ with $i=1,2,3$ form the band with one excitation, $n=1$, and are linear combinations of $\ket{100}$, $\ket{010}$, and $\ket{001}$, whereas $\ket{i}$ with $i=4,5,6$ form the upper band in the spectrum of Fig.~\ref{fig:spectrum} and are linear combinations of $\ket{110}$, $\ket{101}$, and $\ket{011}$. In Fig.~\ref{fig:termalizacion}\,(b,c), we depict the populations of the states in the bands with $n=1$ and 2, i.e., $\rho_{ii}$ for $i=1,\dots,6$, for initial condition $\ket{0}$. Notice that the bands become populated due to collisions and thermalize at long times, even for $\Gamma=1$.

\begin{figure}
    \centering
    \includegraphics[width=.65 \linewidth]{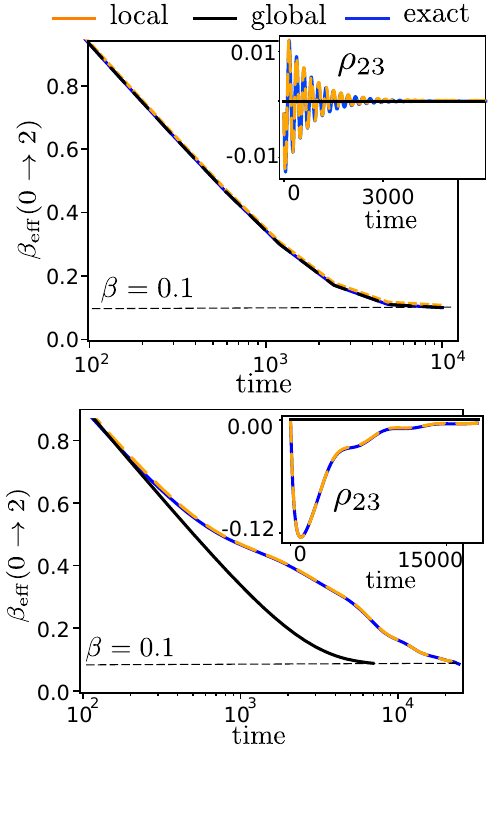}
    \caption{The two plots depict the evolution of the inter-band temperature $\beta_{\rm eff}(0\rightarrow 2)$ and the intra-band coherence $\rho_{23}$ (insets) for $\epsilon=0.1$ (top) and $3\times 10^{-3}$ (bottom). We show the results for the exact evolution, given by the map \eqref{sbig2} (blue), the local limit which neglects terms of order $\Gamma\epsilon$, Eqs.~\eqref{sdecomp1b}--\eqref{sdecomp2bb} (orange), and the global approximation based on wave packets with negligible momentum dispersion, Eq.~\eqref{sbignarrow} (black). The grey line indicates the temperature of the reservoir, which is $\beta = 0.1$.  Other parameters are $m = 0.1$, $g = 50$, $\Gamma = 1$, $\sigma_p = 0.5$ and $h = 4$.}
    \label{fig:temps}
\end{figure}


To explore the thermalization of the system in more detail, we consider the {thermal state} of the chain
\begin{equation}\label{stateglobal}
    \rho_{\rm therm} = \frac{e^{-\beta H}}{Z}
\end{equation}
with partition function $Z = {\rm Tr}\left[e^{-\beta H}\right]$, and the state that results from the thermalization of the first qubit, which we call \textit{local thermal state}:
\begin{equation}\label{statelocal}
    \rho_{\rm loc} = \frac{e^{-\beta h}}{Z_{\rm loc}} \ket{100}\bra{100} + \frac{1}{Z_{\rm loc}} \ket{000}\bra{000}
\end{equation}
with $Z_{\rm loc} = e^{-\beta h} + 1$. Notice that this is the stationary state for $\epsilon = 0$ for initial condition $\ket{0}$.

In Fig.~\ref{fig:termalizacion}\,(e) we plot the population of the first band 
\begin{equation}
  \rho_{123}\equiv \rho_{11}+\rho_{22}+\rho_{33}  
\end{equation}
relative to the population of the ground state $\rho_{00}$, as a function of time for several values of $\epsilon$. The grey lines represent the ratios corresponding to $\rho_{\rm loc}$ and $\rho_{\rm therm}$. An approach to the local thermal state before full thermalization can be appreciated only for very low values of the interaction between qubits, namely $\epsilon\lesssim 10^{-3}$. 
Interestingly, up to a time $\sim 10^3$, trajectories are almost independent of $\epsilon$. The reason is that, for $\Gamma=1$, the evolution at short times is dominated by collisions, i.e., by local transitions between $\ket{000}$ and $\ket{100}$. 



In Figs.~\ref{fig:temps} and \ref{fig:temps2}, we compare the local and global limits with the exact dynamics.
In the former, we plot the interband temperature $\beta_{\rm eff}(0\rightarrow 2)$ for $\epsilon=0.1$ and $3\times 10^{-3}$, where
\begin{equation}
    \beta_{\rm eff}(j\rightarrow k) \equiv \frac{1}{\Delta_{kj}}\log\left(\frac{\rho_{jj}}{\rho_{kk}}\right).
\end{equation}

\begin{figure}
    \centering
    \includegraphics[width=.65\linewidth]{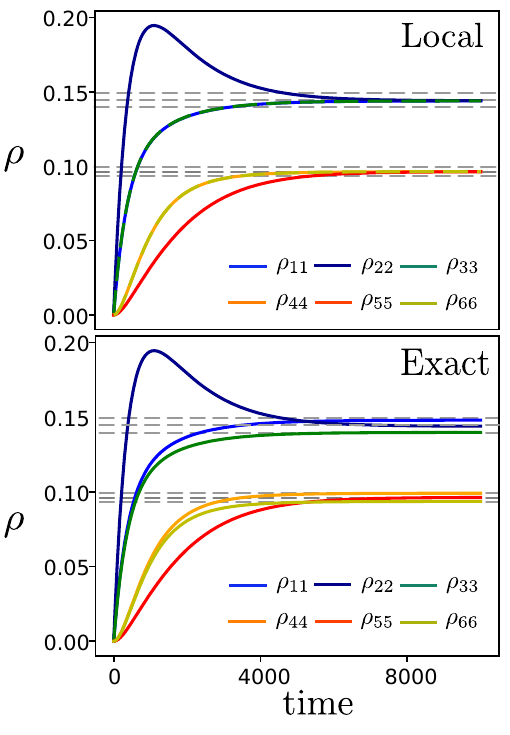}
    \caption{Populations of the eigenstates of the chain Hamiltonian as given by the exact map Eq.~\eqref{sbig2} (bottom panel) and the local limit Eqs.~\eqref{sdecomp1b}--\eqref{sdecomp2bb} (top panel). Observe that, in the local limit, the energy levels within the two bands do not thermalize, as expected.
    In this figure $\Gamma = 1$, with $\epsilon = 0.1$. Other parameters are $\beta = m = 0.1$, $g = 50$, $\sigma_p = 0.5$ and $h = 4$.}
    \label{fig:temps2}
\end{figure}

 As expected, bands thermalize under the three collisional maps, but the global equation fails at short times if $\epsilon$ is small. The local limit, on the other hand, reproduces with high accuracy the evolution of the band population.

Regarding the intra-band coherence $\rho_{23}$, the global equation is unable to create these coherences if the initial state is diagonal in the eigenbasis of the Hamiltonian $H$, which is our case, since $\rho(t=0)=\ket{000}\bra{000}$. This can be seen in the insets of Fig.~\ref{fig:termalizacion}, where the black lines are horizontal. Surprisingly, the local limit reproduces the behavior of the intra-band coherence, $\rho_{23}$, as shown in the insets. The reason is that the stationary state in the local limit is the identity matrix within each band. Therefore, in this limit the intra-band coherences vanish, regardless of whether they are local or global.


Nevertheless, the local limit, as expected,  does not induce the intraband thermalization of the populations. This is shown in Fig.~\ref{fig:temps2}, where the populations of states $\ket{i}$ with $i=1,\dots,6$ are depicted.



\section{Conclusions}
\label{sec:conclusions}

The Lindblad equation derived under weak-coupling and the secular approximation becomes problematic when applied to systems that interact locally with a thermal bath. The resulting equation remains global even when the interaction between the subsystems tends to zero (the limit $\epsilon \to 0$ in the systems analyzed in Sec.~\ref{sec:example}). This occurs because, in this limit, the energy levels become degenerate and the secular approximation is no longer valid.

To overcome this limitation, several works (see~\cite{Bulnes_Cuetara2016-oj, Cattaneo_2019, potts_thermodynamically_2021, PhysRevA.103.062226, PhysRevLett.122.150603,chenEfficientQuantumThermal2025}) have adopted a partial secular approximation, in which nearly degenerate transition frequencies are grouped together. The resulting master equation provides a more accurate description of the system dynamics within its regime of validity than the fully secular one. The thermodynamic consistency of this approach has been examined in~\cite{PhysRevA.103.062226,potts_thermodynamically_2021, Cresser2021Dec}, where it is argued that, for a system coupled to a single thermal bath, the steady state is  $\propto e^{-\beta H^*}$ with $H^*$ differing from the physical Hamiltonian by terms of the order of the Bohr frequency splittings neglected in the partial secular approximation.

In this work, we have used a thermodynamically consistent collisional reservoir to investigate this issue. The resulting dynamics is governed by the Lindblad equation~\eqref{lindblad0}, whose dissipator is given by the collisional map~\eqref{sbig2}. We have shown that this map converges to a local dissipator in the regime of weak coupling between qubits in a chain, ${\Gamma}\color{black} \epsilon \to 0$ , and to a global dissipator when the system is bombarded by particles from a bath with vanishing momentum dispersion, $\sigma_p \to 0$. In our framework, the conflict between local and global Lindblad equations is clearly identified by the noncommutativity of the limits $\epsilon \to 0$ and $\sigma_p \to 0$. The latter yields a global equation that remains global even for $\epsilon \to 0$, whereas the former produces a local equation for any value of $\sigma_p$.

The exact map~\eqref{sbig2} captures both local and global behaviors, depending on the relative magnitudes of the two key parameters, $\epsilon$ and $\sigma_p$. 
Numerical simulations further show that, for finite $\epsilon$ and $\sigma_p$, the exact map accurately reproduces the system's short-time dynamics, while thermalization is approached at long times. Within this regime, one requires a large $\Gamma > \epsilon$ in order to observe local effects during the transient, so that the dynamics is dominated by the collisions. 

{Numerical simulations further show that the local limit accurately reproduces the system’s short-time dynamics, while thermalization is achieved at long times for sufficiently small bombardment rates $\Gamma$\cite{Tabanera-Bravo2023May}, as predicted by the global Lindblad equation. To our knowledge, the collisional map \eqref{sbig2} is the first dissipator that simultaneously predicts the Gibbs state with the actual system Hamiltonian and is local at short times. Hence, it may prove useful for the study of heat transport in quantum systems.

It should be noted, however, that the thermodynamic consistency of the collisional scheme relies on the assumption of weak bombardment, i.e., small $\Gamma$, when the incident wave packets have a non-negligible momentum dispersion. This issue has been discussed in detail in Ref.~\cite{Tabanera-Bravo2023May}, where we also showed numerically that thermalization is achieved even for moderate values of $\Gamma$. Nevertheless, this limitation should be kept in mind when applying the collisional scheme to extended systems in more general settings.
}

Concerning the global limit, note that in extended systems the energy levels are so closely spaced that the momentum dispersion $\sigma_p$ cannot be taken strictly to zero and strict thermalization is never reached. This situation is formally equivalent to the breakdown of the secular approximation. Our collisional map~\eqref{sbig2}, together with the weak-bombardment limit $\Gamma \to 0$, resolves this problem in two complementary ways: (i) the map~\eqref{sbig2} preserves coherences but satisfies detailed balance, thereby ensuring the correct thermalization of populations; and (ii) the combination of a small bombardment rate, $\Gamma \to 0$, and unitary evolution for random times between collisions suppresses coherences \cite{Tabanera-Bravo2023May}, leading to full thermalization.


\section{Aknowledgements}

J.T.-B. acknowledges financial support from  Alexander-von-Humboldt Foundation. F.B.
thanks Fondecyt project 1231210. J.M.R.P. acknowledges financial support from
Grant PID2023-147067NB-I00 funded by MCIU/AEI/10.13039/501100011033 and by ERDF/EU.

\appendix

\begin{widetext}

    \vspace{0.3cm}

\section{Matrices of generalized local operators}
\label{sec:proof}

An interesting scenario yielding generalized local operators occurs when the Hamiltonian $H$ has a dominant local part $H_{\rm loc}$ that commutes with $H$:
\begin{equation}
    H=H_{\rm loc}+ H_{\rm int};\qquad [H_{\rm loc},H_{\rm int}]=0.
\end{equation}
Consider the local eigenbasis of $H_{\rm loc}$, $\ket{s_1,\dots,s_N}$, and the eigenbasis of $H$, $H\ket{j}=\epsilon_j\ket{j}$, which is also eigenbasis of $H_{\rm loc}$:
\begin{align}
    H_{\rm loc}\ket{s_1,\dots,s_N}&=\epsilon_{\{s_i\}}^{(0)}\ket{s_1,\dots,s_N} \\
    H_{\rm loc}\ket{j}&=\epsilon_j^{(0)}\ket{j}.
\end{align}
These equations imply
\begin{align}
      f(\epsilon_j^{(0)})\braket{s_1,\dots,s_N|j}&= f(\epsilon_{\{s_i\}}^{(0)})^*\braket{s_1,\dots,s_N|j}
\\
f(\epsilon_j^{(0)})^*\braket{j|s_1,\dots,s_N}&= f(\epsilon_{\{s_i\}}^{(0)})\braket{j|s_1,\dots,s_N}
\end{align}
for any function $f(\cdot)$.

An operator $L$ verifying:
\begin{equation}   \braket{j'|L|j}=\lambda(\epsilon_{j'}^{(0)},\epsilon_j^{(0)})\braket{j'|L^{(1)}\otimes\mathbb{I}\otimes \dots \otimes \mathbb{I}|j}
\end{equation}
is local in the generalized sense given by Eq.~\eqref{condlocal2}. 
The proof is as follows:
\begin{align}\label{condlocal2app}
   & \braket{s_1',\dots,s_N'|L|s_1,\dots s_N}=\sum_{j'j} \braket{s_1',\dots,s_N'|j'}\braket{j|s_1,\dots s_N}\lambda(\epsilon_{j'}^{(0)},\epsilon_j^{(0)})\braket{j'|L^{(1)}\otimes \mathbb{I}\dots\otimes \mathbb{I}|j}
   \nonumber \\ &=\sum_{j'j} \sum_{\{s_i''\}}\sum_{s_1'''}\braket{s_1',\dots,s_N'|j'}\braket{j|s_1,\dots s_N}  \lambda(\epsilon_{j'}^{(0)},\epsilon_j^{(0)})
\braket{j'|s_1'',\dots,s_N''}
\braket{s_1''',s_2'',\dots,s_N''|j}
\braket{s_1''|L^{(1)}|s_1'''}
\nonumber \\ &=
   \lambda(\epsilon_{\{s_i'\}}^{(0)},\epsilon_{\{s_i\}}^{(0)})
   \sum_{j'j} \sum_{\{s_i''\}}\sum_{s_1'''}\braket{s_1',\dots,s_N'|j'}\braket{j|s_1,\dots s_N}  
\braket{j'|s_1'',\dots,s_N''}
\braket{s_1''',s_2'',\dots,s_N''|j}
\braket{s_1''|L^{(1)}|s_1'''}
   \nonumber \\ &=
   \lambda(\epsilon_{\{s_i'\}}^{(0)},\epsilon_{\{s_i\}}^{(0)})\braket{s_1'|L^{(1)}|s_1}\delta_{s'_2,s_2}\dots \delta_{s_N',s_N}
\end{align}
This expression relates the matrices of generalized local operators in the local basis and in the eigenbasis of $H$.
The same results are obtained if $[H_{\rm loc},H_{\rm int}]=o(\epsilon)\neq 0$ up to linear terms on the commutator $\epsilon$.

\section{Kraus operators of the exact scattering map}
\label{app:kraus}

To find the Kraus operators associated with the scattering map, we must diagonalize the incident density matrix, i.e., find a real positive function $\mu(p_0)$ and pure states $\phi(p;p_0)$, such that
\begin{equation}\label{rhoadiag}
    \rho_{\rm U}(p,p'')=\int dp_0\, \mu(p_0)\phi(p;p_0)\phi(p'';p_0)^*.
\end{equation}
The integration variable $p_0$ labels the eigenvalues $\mu(p_0)$ and the eigenstates $\phi(p;p_0)$ of $\rho_{\rm U}$. 
Introducing \eqref{rhoadiag} in \eqref{sbig}, and comparing with \eqref{stensor2} we find the Kraus operators in the Hamiltonian eigenbasis:
\begin{equation}\label{mbig}
  \braket{j'|M_{\alpha,p_0,E_{\rm out}}|j} =  \int_{0}^\infty dp \, \sqrt{\frac{p\,\mu(p_0)}{m}}\phi(p;p_0)\delta\left(\frac{p^2}{2m}-\Delta_{j'j}-E_{\rm out}\right)s^{(\alpha )}_{j'j}\left(\frac{p^2}{2m} + \epsilon_j\right),
\end{equation}
with $E_{\rm out}\geq 0$.
This expression can be further simplified to obtain
\begin{equation}\label{krausS}
    \braket{j'|M_{\alpha,p_0,E_{\rm out}}|j}=    
    \frac{\sqrt{m \,\mu(p_0)}\,\Theta(\Delta_{j'j}+E_{\rm out})}{[2m(\Delta_{j'j}+E_{\rm out})]^{1/4}}\,
    \phi\left(\sqrt{2m(\Delta_{j'j}+E_{\rm out})}\,;\,p_0\right)    \,s^{(\alpha)}_{j'j}(E_{\rm out}+\epsilon_{j'})
\end{equation}
where $\Theta(\cdot)$ is the Heaviside step function.

The Kraus operators depend on three observables: the pure state $p_0$ of the incident particle, the kinetic energy $E_{\rm out}$ of the particle after the collision, and whether it has been reflected ($\alpha=-$) or transmitted ($\alpha=+$). The effect of the collision on the system is equivalent to a non-selective measurement of these three observables. This Kraus decomposition is not unique. 
In fact, any CPTP map admits a representation with at most $d^2$ Kraus operators 
\cite{Nielsen_Chuang_2010}, whereas $M_{p_0,E_{\rm out},\alpha}$ depends on two continuous parameters, $p_0$ and $E_{\rm out}$.

\vspace{0.4cm}

\end{widetext}

\section{Scattering map for $\epsilon=0$}
\label{app:scat}

Here we explicitly calculate the scattering map for the example in the main text of a chain of qubits bombarded by particles with finite momentum dispersion. 
The Hamiltonian $H$ of the system is given by \eqref{eq:chainmiltonian}
and the interaction $H_{\rm US}=g\,\sigma_x\otimes \delta(\hat x)$, Eq.~\eqref{eq:unitmiltonian},  with the reservoir units is localized at the origin $x=0$. 


The localized interaction allows us to derive an exact formal expression for the scattering matrices $s^{(\alpha)}(E)$, as shown in \cite{Jacob2021Apr}. The matrix corresponding to transmitted particles reads \cite{Jacob2021Apr,Tabanera-Bravo2023May}:
\begin{equation}\label{sdeltacom1}
    s^{(+)}(E)=\mathbb{K}^{1/2}\left[\mathbb{I}+\frac{img}{\hbar^2}\mathbb{K}^{-1}\sigma_x\right]^{-1}\mathbb{K}^{-1/2}
\end{equation}
where $\mathbb{K}=\sqrt{2m(E-H)}/\hbar$,
and the one corresponding to reflected particles is
\begin{equation}\label{sdeltacom0}
    s^{(-)}(E)=s^{(+)}(E)-\mathbb{I}
\end{equation}
The matrices $s^{(\pm)}(E)$ are defined for open channels, i.e., for states $\ket{j}$ such that $\epsilon_j>E$.

\subsection{One qubit}

For  a single qubit, the scattering matrix can be calculated explicitly. If $E>h$:
\begin{equation}
   \mathbb{K}= \frac{\sqrt{2m}}{\hbar}\left(\begin{array}{cc}\sqrt{E-h} & 0\\0 &\sqrt{E} \end{array}\right)
\end{equation}
and equations \eqref{sdeltacom1} and \eqref{sdeltacom0} yield
  \begin{align}
  s^{(+)}(E)&= \frac{1}{1+c^2}\left(\begin{array}{cc}1 & -ic\\-ic & 1 \end{array}\right)\\
   s^{(-)}(E)&= \frac{-c}{1+c^2}\left(\begin{array}{cc} c & i\\i & c \end{array}\right)    
  \end{align}
with
\begin{equation}
    c=\frac{g}{\hbar}\sqrt{\frac{m}{2}}\frac{1}{[E(E-h)]^{1/4}}.
\end{equation}
If $E\leq h$, the only open channel is the elastic one when the system is in the ground state. The scattering matrices are
 \begin{equation}
  s_{00}^{(+)}(E)= \frac{1}{1+c^2};\qquad
   s_{00}^{(-)}(E)= -\frac{c^2}{1+c^2}
\end{equation}  
and now $c^2$ is an imaginary number:
\begin{equation}
    c^2=-i\,\frac{g^2}{\hbar^2}\frac{m}{2}\frac{1}{[E(h-E)]^{1/2}}.
\end{equation}

\subsection{$N$ qubits}

If the total energy is high enough:
\begin{equation}
   E> h\sum_{i=2}^Ns_i +h
\end{equation}
then, for the incident plane wave
\begin{equation}
    e^{-ik_{\{s_i\}}x}\ket{s_1\,s_2\dots s_N}
\end{equation}
with $k_{\{s_i\}}=\sqrt{2m\left(E-h\sum_i s_i\right)}/\hbar$, there are two transmitted waves 
\begin{align}
t_{s_1 s_1}  & e^{-ik_{\{s_i\}}x}\ket{s_1\,s_2\dots s_N}\label{transelas} \\
    t_{\bar s_1 s_1} & e^{-ik_{\{\bar s_1 s_2\dots s_N\}}x}\ket{\bar s_1\,s_2\dots s_N}\label{transnoelas}
\end{align}
where $\bar s_1 = 1- s_1$. The transmitted amplitudes are
\begin{align}
    t_{11} &=t_{00}=\frac{1}{1+c^2} \\
    t_{01}& =t_{10}=\sqrt{\frac{k_{\{s_i\}}}{k_{\{\bar s_1 s_2\dots s_N\}}}}\frac{-ic}{1+c^2}
    \end{align}
with
\begin{equation}\label{cks}
    c=\frac{mg}{\hbar^2}\frac{1}{\sqrt{k_{\{s_i\}}k_{\{\bar s_1 s_2\dots s_N\}}}}.
\end{equation}

If $s_1=0$ and the total energy is small:
\begin{equation}
     h\sum_{i=2}^Ns_i \leq E \leq h\sum_{i=2}^Ns_i +h
\end{equation}
then the only transmitted wave is proportional to $\ket{0\, s_2\dots s_N}$ with amplitude
\begin{equation}
   t_{00}=\frac{1}{1+c^2} 
\end{equation}
and now $c^2$ is an imaginary number, since it is proportional to $1/k_{\{1\, s_2\dots s_N\}}$, which is imaginary.

The eigenstates of the Hamiltonian $H$ can be written in the local basis as:
\begin{equation}\label{jloc}
    \ket{j}=\sum_{\{s_i\}}\braket{s_1\dots s_N|j}\ket{s_1\dots s_N}.
\end{equation}
If $[H_{\rm loc},H_{\rm int}]=0$, then the eigenstates $\ket{j}$ have a well-defined number of excitations $n_j$, and the sum in \eqref{jloc} is restricted to local states with $\sum_i s_i=n_j$ excitations.
An incident wave of the form
\begin{equation}
    e^{-ik_jx}\ket{j}
\end{equation}
with $k_j=\sqrt{2m(E-\epsilon_j)}/\hbar$, can be decomposed as
\begin{equation}\label{wavein}
    \sum_{\{s_i\}}\braket{s_1\dots s_N|j}  e^{-ik_{\{s_i\}}x}\ket{s_1\,s_2\dots s_N}
\end{equation}
Recall that this sum is restricted to local states with $\sum_i s_i=n_j$ excitations. The transmitted waves are of the form \eqref{transelas}-\eqref{transnoelas}, which can be written in the eigenbasis of the Hamiltonian as:
\begin{align}
\sum_k t_{s_1 s_1}  & e^{-ik_{\{s_i\}}x}\braket{k|s_1\,s_2\dots s_N}\ket{k}\label{transelas2} \\
 \sum_k   t_{\bar s_1 s_1} & e^{-ik_{\{\bar s_1 s_2\dots s_N\}}x}\braket{k|\bar s_1\,s_2\dots s_N}\ket{k}.\label{transnoelas2}
\end{align}

\begin{widetext}

Combining \eqref{wavein} with \eqref{transelas2}-\eqref{transnoelas2}, we obtain the following transmision amplitudes for the transition $\ket{j}\to\ket{k}$ through transmitted waves:
\begin{align}
    t_{kj}&= \sum_{\{s_i\}}\braket{s_1\dots s_N|j} \left[t_{s_1 s_1} \braket{k|s_1\,s_2\dots s_N}+t_{\bar s_1 s_1}\braket{k|\bar s_1\,s_2\dots s_N}\right]\nonumber \\
    &=\braket{k|\left(\begin{array}{cc}t_{11} & t_{10} \\t_{01} & t_{00}\end{array}\right)\otimes \mathbb{I}\otimes\dots\otimes \mathbb{I}|j} 
\end{align}

However, now the parameter $c$ depends on the first qubit $s_1$ in the local states of Eq.~\eqref{wavein}, which  can be either 0 or 1. The parameter $c$ in \eqref{cks} contains the wave vector of the ingoing and outgoing plane waves for the non-elastic channel. The first is $\sqrt{2m(E-\epsilon_j)}$, but the second is $\sqrt{2m(E-\epsilon_j)-h}$ if $s_1=0$ and $\sqrt{2m(E-\epsilon_j)+h}$ if $s_1=1$. Hence, the transmitted amplitudes read
\begin{equation}
     t_{kj}=\braket{k|\left(\begin{array}{cc} \displaystyle\frac{1}{1+c_+^2} &\displaystyle \left(\frac{E-\epsilon_j}{E-\epsilon_j-h}\right)^{1/4}\frac{-ic_-}{1+c_-^2} \\[0.4cm] \displaystyle \left(\frac{E-\epsilon_j}{E-\epsilon_j+h}\right)^{1/4}\frac{-ic_+}{1+c_+^2}& \displaystyle \frac{1}{1+c_-^2}\end{array}\right)\otimes \mathbb{I}\otimes\dots\otimes \mathbb{I}|j}
\end{equation}
with
\begin{equation}
    c_\pm=\frac{g}{\hbar}\sqrt{\frac{m}{2}}\frac{1}{\left[(E-\epsilon_j)(E-\epsilon_j\pm h)\right]^{1/4}}.
\end{equation}

Consider now the case of small energy, obeying $E<\epsilon_j+h$. Again, the first qubit $s_1$ can be either 0 or 1 in the local states of Eq.~\eqref{wavein}. If $s_1=1$, the elastic and inelastic channels are open, whereas if $s_1=0$, only the elastic channel is open. Hence, the transmitted amplitudes are
\begin{equation}
     t_{kj}=\braket{k|\left(\begin{array}{ccc} \displaystyle\frac{1}{1+c_+^2} &\qquad & 0 \\[0.4cm] \displaystyle \left(\frac{E-\epsilon_j}{E-\epsilon_j+h}\right)^{1/4}\frac{-ic_+}{1+c_+^2}&  \qquad \qquad&\displaystyle \frac{1}{1+c_-^2}\end{array}\right)\otimes \mathbb{I}\otimes\dots\otimes \mathbb{I}|j}
\end{equation}
and now $c_-^2$ is an imaginary number.

Finally, the scattering matrices are:
\begin{align}
    s^{(+)}_{kj}(E)&=\braket{k|\left(\begin{array}{cc} \displaystyle\frac{1}{1+c_+^2} &\displaystyle \frac{-ic_-}{1+c_-^2} \\[0.4cm] \displaystyle \frac{-ic_+}{1+c_+^2}& \displaystyle \frac{1}{1+c_-^2}\end{array}\right)\otimes \mathbb{I}\otimes\dots\otimes \mathbb{I}|j},
    \qquad s^{(-)}_{kj}(E)=s^{(+)}_{kj}(E)-\delta_{kj} 
    \qquad \mbox{if $E>\epsilon_j+h$}\\[0.4cm]
     s^{(+)}_{kj}(E)&=\braket{k|\left(\begin{array}{cc} \displaystyle\frac{1}{1+c_+^2} & 0  \\[0.4cm] \displaystyle \frac{-ic_+}{1+c_+^2}& \displaystyle \frac{1}{1+c_-^2}\end{array}\right)\otimes \mathbb{I}\otimes\dots\otimes \mathbb{I}|j},
    \qquad s^{(-)}_{kj}(E)=s^{(+)}_{kj}(E)-\delta_{kj} 
    \qquad \mbox{if $\epsilon_j<E<\epsilon_j+h$}\\
\end{align}

The matrices that appear in the scattering map \eqref{sdecomp1}-\eqref{sdecomp2} can be written as:
\begin{align}
    s^{(+)}_{kj}(E_p+\epsilon_j)&=\braket{k|\left(\begin{array}{cc} \displaystyle\frac{1}{1+c_+(p)^2} &\displaystyle \frac{-ic_-(p)}{1+c_-(p)^2} \\[0.4cm] \displaystyle \frac{-ic_+}{1+c_+(p)^2}& \displaystyle \frac{1}{1+c_-(p)^2}\end{array}\right)\otimes \mathbb{I}\otimes\dots\otimes \mathbb{I}|j}
    \qquad \mbox{if $E>\epsilon_j+h$}\\
     s^{(+)}_{kj}(E_p+\epsilon_j)&=\braket{k|\left(\begin{array}{cc} \displaystyle\frac{1}{1+c_+(p)^2} & 0 \\[0.4cm] \displaystyle \frac{-ic_+(p)}{1+c_+(p)^2}& \displaystyle \frac{1}{1+c_-(p)^2}\end{array}\right)\otimes \mathbb{I}\otimes\dots\otimes \mathbb{I}|j}
    \qquad \mbox{if $\epsilon_j<E<\epsilon_j+h$}
\end{align}
with
\begin{equation}
    c_\pm(p)=\frac{gm}{\hbar}\frac{1}{\left[p\sqrt{p^2\pm 2mh}\right]^{1/2}}
\end{equation}
which verify:
\begin{equation}
    \int_0^\infty dp\, \mu_{\rm eff}(p)f(c_+(p))= e^{\beta h}\int_{\sqrt{2mh}}^\infty dp\, \mu_{\rm eff}(p)f(c_-(p))
\end{equation}
for any function $f(\cdot)$. This identity results from the change of variable $p'=\sqrt{p^2+2mh}$ in the integral of the l.h.s.

\subsection{Coherences}

Let us calculate the first-order correction $\rho^{(1)}$ from Eqs.~\eqref{pert1} and \eqref{sum}. We first calculate $\sum_j \mathbb{S}^{jj}_{j'k'}$ when the three states $j,j',k'$ are in the same band. The contribution of the transmitted waves is:
\begin{align}
    &\sum_j \int_0^\infty dp\,\mu_{\rm eff}(p)
    \braket{j'|\left(\begin{array}{cc} \displaystyle\frac{1}{1+c_+(p)^2} & 0 \\ 0 & \displaystyle \frac{1}{1+c_-(p)^2}\end{array}\right)\otimes \mathbb{I}\otimes\dots\otimes \mathbb{I}|j}\braket{k'|\left(\begin{array}{cc} \displaystyle\frac{1}{1+c_+(p)^2} & 0 \\ 0 & \displaystyle \frac{1}{1+c_-(p)^{2}}\end{array}\right)\otimes \mathbb{I}\otimes\dots\otimes \mathbb{I}|j}* \nonumber \\
    =&\int_0^\infty dp\,\mu_{\rm eff}(p) \braket{j'|\left(\begin{array}{cc} \displaystyle\frac{1}{(1+c_+(p)^2)^2} & 0 \\ 0 & \displaystyle \frac{1}{|1+c_-(p)^2|^2}\end{array}\right)\otimes \mathbb{I}\otimes\dots\otimes \mathbb{I}|k'}
\end{align}
Now, we calculate the sum when $\epsilon_{j'}=\epsilon_{k'}=\epsilon_j+h$:
\begin{align}
    &\sum_j \int_{\sqrt{2mh}}^\infty dp\,\mu_{\rm eff}(p)
    \braket{j'|\left(\begin{array}{cc} 0 & \displaystyle\frac{-ic_-(p)}{1+c_-(p)^2}  \\ 0 & 0 \end{array}\right)\otimes \mathbb{I}\otimes\dots\otimes \mathbb{I}|j}\braket{k'|\left(\begin{array}{cc} 0 &\displaystyle\frac{-ic_-(p)}{1+c_-(p)^2} \\ 0 & 0 \end{array}\right)\otimes \mathbb{I}\otimes\dots\otimes \mathbb{I}|j}* \nonumber \\
    =&\sum_j \int_{\sqrt{2mh}}^\infty dp\,\mu_{\rm eff}(p)
    \braket{j'|\left(\begin{array}{cc} 0 & \displaystyle\frac{-ic_-(p)}{1+c_-(p)^2}  \\ 0 & 0 \end{array}\right)\otimes \mathbb{I}\otimes\dots\otimes \mathbb{I}|j}\braket{j|\left(\begin{array}{cc} 0 & 0\\ \displaystyle\frac{ic_-(p)}{1+c_-(p)^2} & 0  \end{array}\right)\otimes \mathbb{I}\otimes\dots\otimes \mathbb{I}|k'} \nonumber \\
    =&\int_{\sqrt{2mh}}^\infty dp\,\mu_{\rm eff}(p)\braket{j'|\left(\begin{array}{cc} \displaystyle\frac{c_-(p)^2}{(1+c_-(p)^2)^2} & 0 \\ 0 &  0 \end{array}\right)\otimes \mathbb{I}\otimes\dots\otimes \mathbb{I}|k'}
    \nonumber \\
    =&\,e^{-\beta h}\int_0^\infty dp\,\mu_{\rm eff}(p)\braket{j'|\left(\begin{array}{cc} \displaystyle\frac{c_+(p)^2}{(1+c_+(p)^2)^2} & 0 \\ 0 &  0 \end{array}\right)\otimes \mathbb{I}\otimes\dots\otimes \mathbb{I}|k'}
\end{align}
The case $\epsilon_{j'}=\epsilon_{k'}=\epsilon_j-h$ reads
\begin{align}
    &\sum_j \int_0^\infty dp\,\mu_{\rm eff}(p)
    \braket{j'|\left(\begin{array}{cc} 0 & 0 \\ \displaystyle\frac{-ic_+(p)}{1+c_+(p)^2}   & 0 \end{array}\right)\otimes \mathbb{I}\otimes\dots\otimes \mathbb{I}|j}\braket{k'|\left(\begin{array}{cc} 0 & 0 \\ \displaystyle\frac{-ic_+(p)}{1+c_+(p)^2}  & 0 \end{array}\right)\otimes \mathbb{I}\otimes\dots\otimes \mathbb{I}|j}* \nonumber \\
    =&\sum_j \int_{0}^\infty dp\,\mu_{\rm eff}(p)
    \braket{j'|\left(\begin{array}{cc} 0 & 0 \\ \displaystyle\frac{-ic_+(p)}{1+c_+(p)^2}   & 0 \end{array}\right)\otimes \mathbb{I}\otimes\dots\otimes \mathbb{I}|j}\braket{j|\left(\begin{array}{cc} 0 &  \displaystyle\frac{ic_+(p)}{1+c_+(p)^2}\\ 0 & 0  \end{array}\right)\otimes \mathbb{I}\otimes\dots\otimes \mathbb{I}|k'} \nonumber \\
    =&\int_0^\infty dp\,\mu_{\rm eff}(p)\braket{j'|\left(\begin{array}{cc} 0 & 0 \\ 0 & \displaystyle\frac{c_+(p)^2}{(1+c_+(p)^2)^2}  \end{array}\right)\otimes \mathbb{I}\otimes\dots\otimes \mathbb{I}|k'}
    \nonumber \\
    =&\,e^{\beta h}\int_{\sqrt{2mh}}^\infty dp\,\mu_{\rm eff}(p)\braket{j'|\left(\begin{array}{cc} 0 & 0 \\ 0 & \displaystyle\frac{c_-(p)^2}{(1+c_-(p)^2)^2}  \end{array}\right)\otimes \mathbb{I}\otimes\dots\otimes \mathbb{I}|k'}
\end{align}
Setting the zero energy at $\epsilon_{j'}=\epsilon_{k'}$ and considering only transmitted waves, we have:
\begin{align}
    (+):\sum_{j}\mathbb{S}^{jj}_{j'k'}e^{-\beta\epsilon_j} &=
 \braket{j'|\left(\begin{array}{cc} \displaystyle\int_0^\infty dp\,\mu_{\rm eff}(p)\frac{1}{1+c_+(p)^2} & 0 \\ 0 & \displaystyle\int_{\sqrt{2mh}}^\infty dp\,\mu_{\rm eff}(p)\frac{1}{1+c_-(p)^2}\end{array}\right)\otimes \mathbb{I}\otimes\dots\otimes \mathbb{I}|k'}\nonumber \\ &+
 \displaystyle\int_0^{\sqrt{2mh}}dp\,\mu_{\rm eff}(p)\braket{j'|\left(\begin{array}{cc} 0 & 0 \\ 0 & \displaystyle \frac{1}{|1+c_-(p)^2|^2}\end{array}\right)\otimes \mathbb{I}\otimes\dots\otimes \mathbb{I}|k'}
\end{align}
Proceeding in the same way for reflected waves, we get:
\begin{align}
  (-):  \sum_{j}\mathbb{S}^{jj}_{j'k'}e^{-\beta\epsilon_j} &=
 \braket{j'|\left(\begin{array}{cc} \displaystyle\int_0^\infty dp\,\mu_{\rm eff}(p)\frac{c_+(p)^2+c_+(p)^4}{(1+c_+(p)^2)^2} & 0 \\ 0 & \displaystyle\int_{\sqrt{2mh}}^\infty dp\,\mu_{\rm eff}(p)\frac{c_-(p)^2+c_-(p)^4}{(1+c_-(p)^2)^2}\end{array}\right)\otimes \mathbb{I}\otimes\dots\otimes \mathbb{I}|k'}\nonumber \\ &+
 \int_0^{\sqrt{2mh}}dp\,\mu_{\rm eff}(p)\braket{j'|\left(\begin{array}{cc} 0 & 0 \\ 0 & \displaystyle \frac{|c_-(p)|^4}{|1+c_-(p)^2|^2}\end{array}\right)\otimes \mathbb{I}\otimes\dots\otimes \mathbb{I}|k'}
\end{align}

Finally, summing all terms, we obtain:
\begin{equation}\label{complete}
    \sum_{j}\mathbb{S}^{jj}_{j'k'}\frac{e^{-\beta\epsilon_j}}{Z}=\delta_{j'k'}\,\frac{e^{-\beta\epsilon_{j'}}}{Z}
\end{equation}

\end{widetext}

\bibliography{bibliography}

\end{document}